\documentclass[aps,prl,twocolumn,superscriptaddress]{revtex4}
\usepackage{float}
\usepackage{graphicx}
\usepackage{dcolumn}
\usepackage{bm}
\usepackage{color}
\usepackage{float}
\usepackage{amsmath}
\usepackage{amssymb}
\usepackage[final]{pdfpages}

\newcommand{\eph}{{\it e}-ph}

\allowdisplaybreaks

\begin{document} 
\title{Quantum Monte Carlo study of lattice polarons in the two-dimensional multi-orbital 
Su-Schrieffer-Heeger model}
\author{Shaozhi Li} 
\affiliation{Department of Physics and Astronomy, The
University of Tennessee, Knoxville, Tennessee 37996, USA} 
\affiliation{Department of Physics, University of Michigan, Ann Arbor, Michigan 48109, USA}
\author{Steven Johnston} 
\email{sjohn145@utk.edu}
\affiliation{Department of Physics and Astronomy, The University of
Tennessee, Knoxville, Tennessee 37996, USA} 
\affiliation{Joint Institute for
Advanced Materials at The University of Tennessee, Knoxville, Tennessee 37996, USA} 
\date{\today}

\begin{abstract}
We study a three-orbital Su-Schrieffer-Heeger model defined on a two-dimensional 
Lieb lattice and in the negative charge transfer regime using determinant quantum Monte Carlo. 
At half-filling (1 hole/unit cell), we observe a bipolaron insulating phase, 
where the ligand oxygen atoms collapse and expand about alternating
cation atoms to produce a bond-disproportionated state. This phase 
is robust against moderate hole doping but is eventually suppressed at large
hole concentrations, leading to a metallic polaron-liquid-like state with
fluctuating patches of local distortions.  
Our results suggest that
the polarons are highly disordered in the metallic state and freeze into a
periodic array across the metal-to-insulator transition. 
We also find an $s$-wave superconducting state at finite doping that primarily appears on the oxygen 
sublattices. Our approach provides an 
efficient, non-perturbative way to treat bond phonons in higher dimensions 
and our results have implications for many materials where coupling to 
bond phonons is the dominant interaction. 
\end{abstract}

\maketitle
{\it Introduction} --- 
Model Hamiltonians for electron-phonon ({\eph}) interactions are broadly divided 
into two categories based on whether the coupling is diagonal or off-diagonal 
in orbital space. Diagonal {\eph} interactions ({\it e.g.}, Holstein \cite{Holstein} 
or Fr{\"o}hlich \cite{Frohlich} models) couple the
atomic displacements to the charge density while 
off-diagonal {\eph} couplings ({\it e.g.}, the 
Su-Schrieffer-Heeger [SSH] model\cite{SuPRL1979}) modulate 
the carrier's kinetic energy via the overlap integrals.  
To date, diagonal {\eph} interactions have received the most 
attention \cite{Sangiovanni, Macridin, Bauer2010, BauerPRB,
Nowadnick, MurakamiPRB, ShaozhiPRB2017, BerciuPRL,NoackPRL,Costa2018,
Bonca, KuPRB2002,HaguePRB2006,
RomeroPRB,shaozhiPRB2015,Chakraverty,Diagrams,HohenadlerPRB2018,DeePRB2019},  while 
studies of off-diagonal models have mainly been restricted to one-dimension
[1D]~\cite{SuPRL1979, Shaozhi2013,Marchand,Sous,SenguptaPRB, Hohenadler2016, TangPRB1988}. 

There is an urgent need to address off-diagonal {\eph} interactions in  
higher dimensions, 
because such couplings are not only relevant to many materials -- {\it e.g.} the organic
charge-transfer solids \cite{SuPRL1979, Shaozhi2013, Clay2018}, 
the rare-earth nickelates \cite{Medarde, Shamblin2018,
JohnstonPRL}, and high-$\mathrm{T}_\mathrm{c}$ superconductors like the cuprates
\cite{Lanzara_Nature, Weber_PRL} and bismuthates \cite{KhazraiePRB} --  
but several recent studies in the few-particle limit have
shown that the new physics can occur in such models. 
For example, strong off-diagonal interactions can produce highly mobile
polarons with light effective masses \cite{Marchand}, generate robust
phonon-mediated pairing \cite{Sous}, and even stabilize and control the
location of a type-II Dirac point \cite{Moeller2017}. 
Off-diagonal models have also gained attention in relation to 1D 
topological insulators \cite{QiRMP2011} in the BDI class \cite{SchnyderPRB2008}.
It is, therefore, imperative to study off-diagonal {\eph} interactions in higher dimensions and at arbitrary fillings, as our intuition gained by studying diagonal models may not serve us well.  

Another motivation for studying the SSH-type models in higher dimensions is to better 
understand its role in establishing the insulating and superconducting states 
of the high-T$_c$ superconducting bismuthates $\mathrm{Ba}_{1-x}\mathrm{K}_x\mathrm{BiO}_3$ (BKBO).
BKBO is in the so-called ``negative charge transfer'' regime \cite{Mizokawa, ZSA,
Foyevtsova_PRB, Plumb2016}, where holes self-dope from the cation to the ligand
oxygen atoms. The subsequent hybridization between the cation and the oxygen atoms then leads to a sizable {\eph} interaction \cite{JohnstonPRL,KhazraiePRB}, 
which may be further enhanced by correlations \cite{PhysRevX.3.021011}, and is believed to drive a high-temperature metal-to-insulator (MIT) transition. 
Here, the insulating state
has a bond disproportionated structure with expanded and collapsed BiO$_6$
octahedra alternating through the material and pairs of holes
condensed into the molecular orbitals formed from the ligand oxygen 
orbitals with $A_{1g}$ symmetry \cite{Foyevtsova_PRB,
Plumb2016, KhazraiePRB, JohnstonPRL, Park2012, Bisogni2016}. 
The relevant model describing this case is a multiorbital SSH model; however, 
knowledge about such models is limited due to a lack of suitable approaches 
for solving it.

With this motivation, we developed a determinant quantum Monte Carlo (DQMC) method for simulating SSH-type interactions, which is applied to study a 2D multi-orbital model for the first time. At half-filling (one hole/unit cell), we find
that the system is a bipolaronic insulator with a bond-disproportionated 
structure, similar to what is observed in BKBO \cite{Sleight} or the  
rare-earth nickelates \cite{Medarde}.  
Hole doping suppresses the insulating phase, giving way to state where the lattice distortions have short-range correlations suggestive of phase mixing and/or fluctuations. 
At high doping levels, we find evidence for a metallic phase where holes are strongly correlated with local structural distortions, forming a polaron-liquid phase. 
Finally, at low temperatures, we find $s$-wave superconducting tendencies that form primarily on the oxygen sublattice and evidence for a superconducting dome. 
Our results are in qualitative agreement with the phase diagram of the bismuthate superconductors and provide theoretical support for a polaronic view of BKBO 
and other negative charge transfer oxides. 

\begin{figure}[t]
\center\includegraphics[width=0.99\columnwidth]{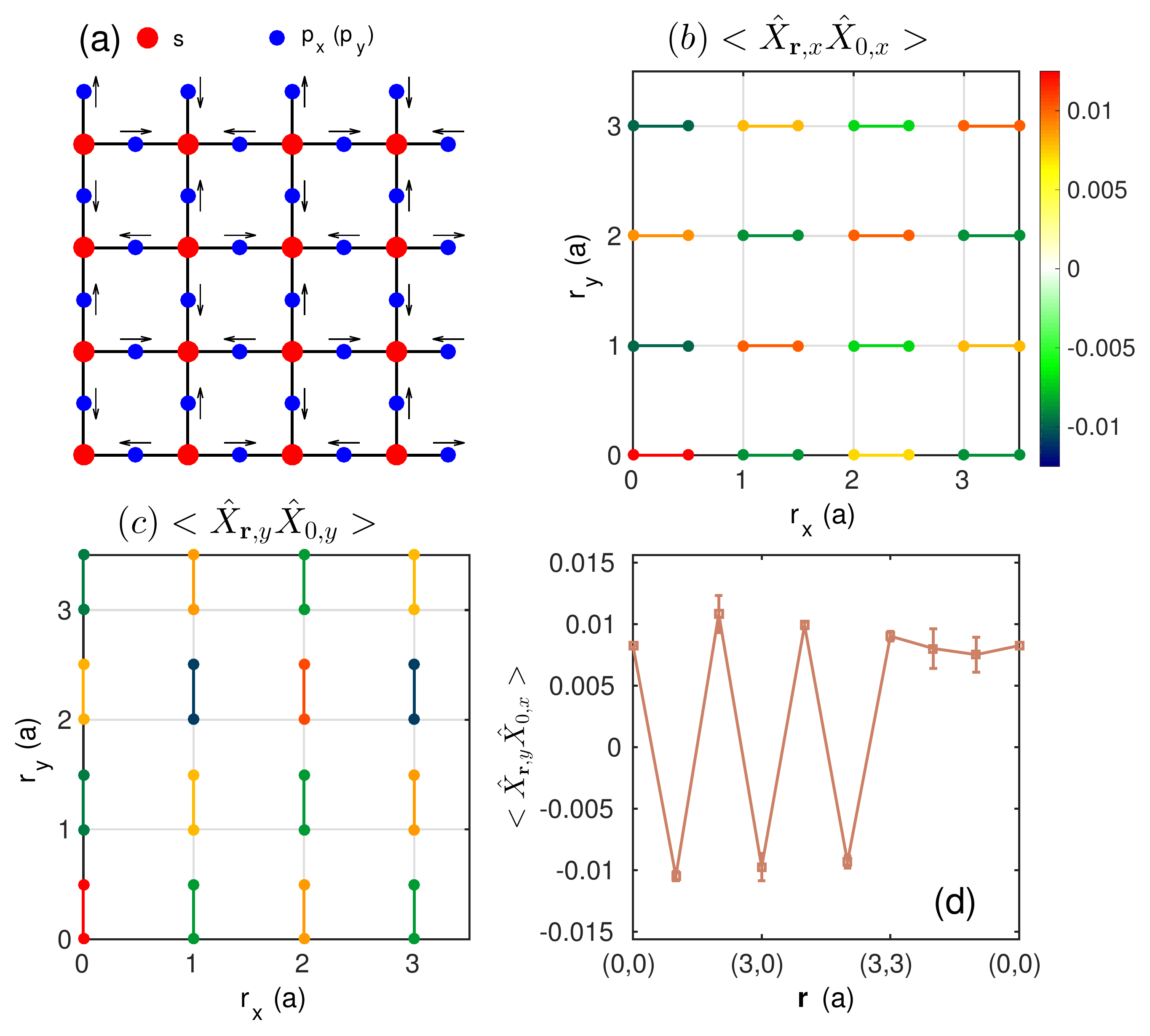}
\caption{\label{Fig:fg1} (a) A sketch of bond disproportionated lattice 
structure. The
red and blue dots indicate the $s$ and $p_{x,y}$ orbitals, respectively, while
the black arrow indicate the displacement pattern of each oxygen atom in the
bond disproportionated structure.  Panels (b) and (c) plot the lattice displacement
correlation functions $\langle \hat{X}_{{\bf r},x} \hat{X}_{0,x}\rangle$ and $\langle
\hat{X}_{{\bf r},y} \hat{X}_{0,y}\rangle$ as a function of distance ${\bf r}=n_x {\bf
a} + n_y {\bf b}$, respectively. Here, ${\bf a}$ and ${\bf b}$ are the
primitive vectors along x- and y-directions, respectively. Panel (d) plots the
real-space displacement correlation function $\langle \hat{X}_{{\bf r},y}
\hat{X}_{0,x}\rangle$ indicating the two-sublattice structure of the  bond disproportionated  
state.  The distance between two nearest Bi atom in the undistorted square
structure is $a$.} \end{figure}

\begin{figure}[t] 
\center\includegraphics[width=0.8\columnwidth]{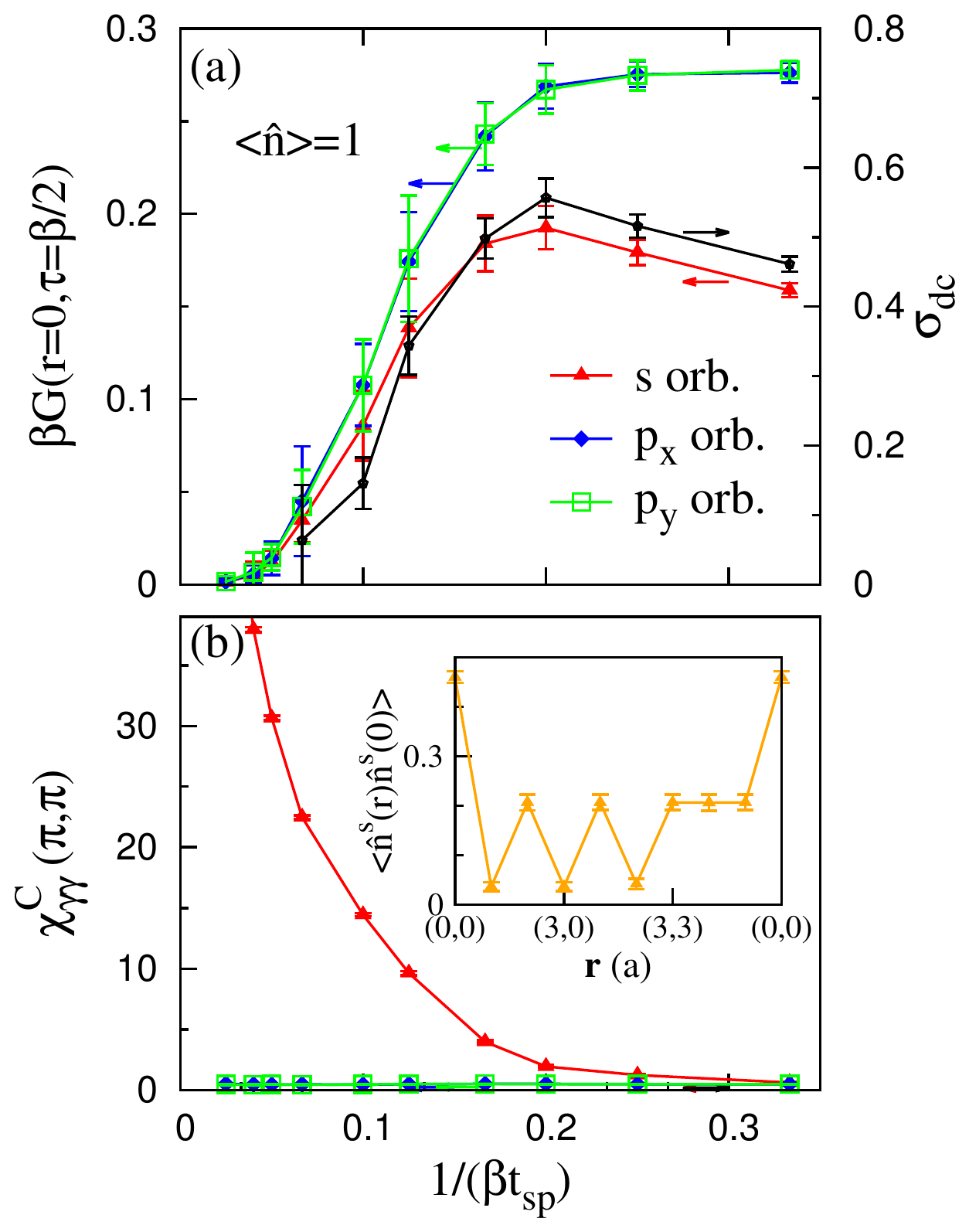}
\caption{\label{Fig:fg2} (a) The temperature dependence of the spectral weight
at the Fermi level $\beta G(r=0,\tau=\beta/2)$ and the direct current (dc)
conductivity $\sigma_\mathrm{dc}$.  (b) The temperature dependence of the
charge-density-wave susceptibility $\chi_C(\pi,\pi)$.  In both
panels, the average filling is $\langle n \rangle =1$ corresponding to the
``half-filled" case with one hole per unit cell. } 
\end{figure}

\begin{figure*}[t] 
\center\includegraphics[width=\textwidth]{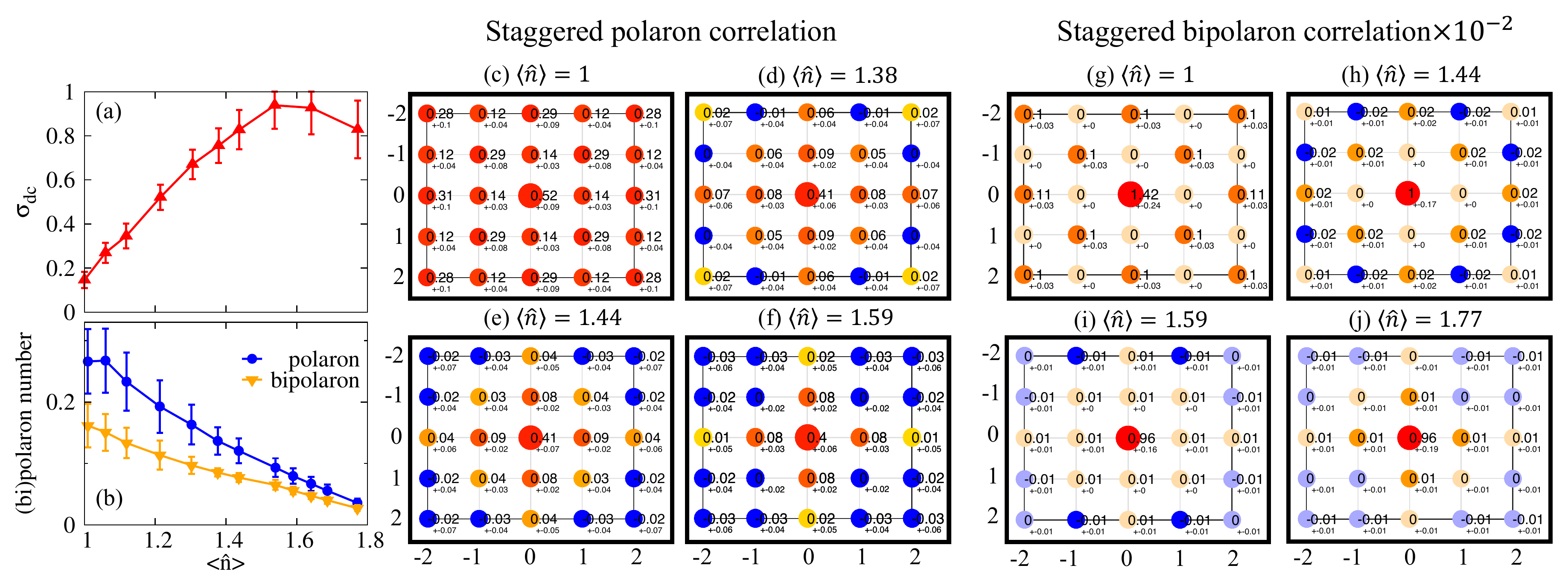}
\caption{\label{Fig:fg3} (a) The dc conductivity as a function of doping. (b)
polaron and bipolaron number as a
function of doping. (c)-(f) Staggered polaron
correlation function $\langle P({\bf r}) \rangle$ and (h)-(j) staggered
bipolaron correlation function $\langle BP({\bf r}) \rangle$ at different
doping levels. The red (gray) color indicates values larger (smaller) than
zero. The numerical value of the correlation function, along with the
associated $1\sigma$ statistical error are indicated at each point. All results
are for a temperature $1/(\beta t_{sp})=0.1$ and error bars smaller than the
marker size have been suppressed for clarity.} \end{figure*}

{\it Model and Methods} --- 
Keeping BKBO in mind, we adopt a three-orbital SSH model defined on a Lieb lattice whose orbital basis consists of a Bi $6s$ orbital and two O $2p$ orbitals, as shown in Fig. \ref{Fig:fg1}(a).  
We freeze the heavier Bi atoms 
into place and restrict lighter O atoms to move along the bond
directions.  The Hamiltonian is $H=H_0+H_\mathrm{lat}+H_{\eph}$, where
\begin{eqnarray}\label{eq:Heph} 
H_0&=&-t_{sp} \sum_{\langle {\bf r},\delta\rangle,\sigma}  \left(
P_{\delta}^{\phantom\dagger} s_{{\bf r},\sigma}^{\dagger}p_{{\bf
r},\delta,\sigma}^{\phantom\dagger} + \mathrm{h.c.} \right) \nonumber\\
&+& t_{pp}\sum_{\langle{\bf
r}, \delta,\delta^\prime\rangle,\sigma} \nonumber
P_{\delta,\delta^\prime}^{\phantom\dagger} p_{{\bf
r},\delta,\sigma}^{\dagger}p_{{\bf r},\delta^\prime,\sigma}^{\phantom\dagger}\nonumber\\
&+&
\sum_{{\bf r},\sigma} \Big[ (\epsilon_s-\mu)\hat{n}^s_{{\bf r},\sigma}
+ (\epsilon_p-\mu)(\hat{n}^{p_x}_{{\bf r},\sigma}
+\hat{n}^{p_y}_{{\bf r},\sigma}) \Big], \nonumber \\ 
H_\mathrm{lat}&=&\sum_{{\bf
r}} \left( \frac{\hat{P}_{{\bf r},x}^2}{2M} + K\hat{X}_{{\bf r},x}^2 +
\frac{\hat{P}_{{\bf r},y}^2}{2M} + K\hat{X}_{{\bf r},y}^2\right) \nonumber\\
H_{\eph}&=& \alpha t_{sp} \sum_{\langle {\bf r},\delta\rangle,\sigma} \left(   \hat{u}_{{\bf
r},\delta}^{\phantom\dagger} s_{{\bf r},\sigma}^{\dagger}p_{{\bf
r},\delta,\sigma}^{\phantom\dagger} + \mathrm{h.c.} \right). \nonumber 
\end{eqnarray}
Here, $\langle \dots \rangle$ denotes a sum over nearest neighbor 
atoms, $\delta, \delta^\prime=\pm x$, $\pm y$ index the oxygen atoms
surrounding each Bi, and the operators $s_{{\bf r},\sigma}^{\dagger}$ 
$\left( s_{{\bf r},\sigma}^{\phantom\dagger}\right)$ and $p_{{\bf
r},\delta,\sigma}^{\dagger}$ $\left(p^{\phantom\dagger}_{{\bf
r},\delta,\sigma}\right)$ create (annihilate) 
spin $\sigma$ holes on the Bi $6s$ and O $2p_\delta$ orbitals, respectively. 
The unit cells are indexed by ${\bf r}= n_x {\bf a} + n_y{\bf b}$, where  
${\bf a} = (a,0)$, ${\bf b} = (0,a)$ are the primitive lattice 
vectors along $x$- and $y$-directions, 
respectively, and $a$ is the Bi-Bi bond length (and our unit of length).   
To simplify the notation, we have introduced shorthand notation 
$p_{{\bf r},-x,\sigma}=p_{{\bf r}-{\bf a},x,\sigma}$ and
$p_{{\bf r},-y,\sigma}=p_{{\bf r}-{\bf b},y,\sigma}$. 
The operators $\hat{n}^s_{{\bf
r},\sigma}=s_{{\bf r},\sigma}^{\dagger}s_{{\bf r},\sigma}^{\phantom\dagger}$
and $\hat{n}^{p_\alpha}_{{\bf r},\sigma}=p_{{\bf
r},\alpha,\sigma}^{\dagger}p_{{\bf r},\alpha,\sigma}^{\phantom\dagger}$ are the 
number operators for $s$ and $p_\alpha$ ($\alpha = x,y$) orbitals,
respectively; $\epsilon_s$ and $\epsilon_p$ are the site energies; $\mu$ is
the chemical potential; $t_{sp}$ and $t_{pp}$ are the Bi-O and
O-O hopping integrals in the undistorted crystal; and $\alpha$ is the {\eph}
coupling constant. The phase factors are $P_{x(y)}=-P_{-x(-y)}=1$, and $P_{\pm
x,\pm y}=P_{\pm y, \pm x}=-P_{\pm x,\mp y}=-P_{\mp y, \pm x}=1$.  The motion of
the O atoms described by the atomic displacement (momentum) operators 
$\hat{X}_{{\bf r}, \alpha}$ ($\hat{P}_{{\bf r}, \alpha}$). Here, $M$ is the oxygen mass and $K$ is the coefficient of
elasticity between each Bi and O atom, and each O is linked by springs 
to the neighboring Bi atoms. Thus, bare phonon frequency is
$\Omega=\sqrt{2K/M}$.  Finally, the atomic displacements modulate the
hopping integral as $t_{sp}(P_{\delta}-\alpha \hat{u}_{{\bf r},\delta})$, where 
we have introduced the shorthand 
$\hat{u}_{{\bf r},x}=\hat{X}_{{\bf r},x}$, $\hat{u}_{{\bf r},-x}=\hat{X}_{{\bf
r}-{\bf a},x}$, $\hat{u}_{{\bf r},y}=\hat{X}_{{\bf r},y}$, and $\hat{u}_{{\bf
r},-y}=\hat{X}_{{\bf r}-{\bf b},y}$. 

We study the model on a square lattice with $N=4\times 4$ Bi atoms (48 orbitals in total) using 
DQMC. We stress that the model considered here is free of the Fermion sign problem. 
The details are provided in the supplementary materials~\cite{Supplement},  
along with expressions for the standard quantities measured in this work, and supplementary exact diagonalization calculations. 
The details of all non-standard quantities are provided in the main text. 
Throughout, we adopt $t_{sp}=2.08$, $t_{pp}=0.056$, $\epsilon_s=6.42$, and
$\epsilon_p=2.42$ (in units of eV), which are obtained from DFT calculations 
of BaBiO$_3$ \cite{KhazraiePRB}. 
We adopt a phonon energy $\Omega=\sqrt{2}t_{sp}$ and {\eph} coupling 
strength $\alpha=4a^{-1}$, which gives average displacement's squared of  
$\frac{1}{N}\sum_{\bf r}\langle \hat{X}^2_{{\bf r},x} \rangle = \frac{1}{N}\sum_{\bf r} \langle \hat{X}^2_{{\bf r},y} \rangle
= 0.0356a^2$ at half-filling, indicating that the oxygen atoms do
not cross the bismuth atoms during the sampling. 
(Here, we are limited to large $\Omega$ by long autocorrelation times.) 

{\it Results} --- 
Figures \ref{Fig:fg1}(b)-\ref{Fig:fg1}(d) plots the 
lattice displacement correlation functions $\langle \hat{X}_{{\bf r},x}
\hat{X}_{0,x}\rangle$, $\langle \hat{X}_{{\bf r},y} \hat{X}_{0,y}\rangle$, and $\langle
\hat{X}_{{\bf r},y} \hat{X}_{0,x}\rangle$, as a function of position at
inverse  temperature $\beta = 10/t_{sp}$, which provides 
evidence for a bond disproportionated insulating state at $\langle \hat{n} \rangle=1$. 
Both $\langle \hat{X}_{{\bf r},x}
\hat{X}_{0,x}\rangle$ and $\langle \hat{X}_{{\bf r},y} \hat{X}_{0,y}\rangle$ alternate
in sign following a checkerboard pattern while $\langle \hat{X}_{{\bf r},y}
\hat{X}_{0,x}\rangle$ alternates in sign along $x$- and $y$-directions but is
constant along the diagonal. This behavior reflects the breathing
distortion sketched in Fig. \ref{Fig:fg1}(a), and is consistent with 
the bond disproportionation observed in the insulating phase of the 
BKBO \cite{Cox1979, Rice, KimPRL2015}. 

Figure \ref{Fig:fg2}(a) plots the dc conductivity $\sigma_\mathrm{dc}$ and orbital-resolved spectral weight 
$\beta G_{\gamma\gamma}({\bf r}=0,\tau=\beta/2)$, where $\gamma$ is the
orbital index, for $\langle \hat{n} \rangle = 1$~\cite{TrivediPRB,Supplement}.  
The conductivity (black dots) initially increases as the 
temperature is lowered until reaching a maximum at $\beta \approx 5/ t_{sp}$ 
then it is suppressed. All three orbital spectral weights 
follow a similar trend, indicating a concomitant removal of spectral weight at the 
Fermi level.    
The insulating phase is characterized by a ${\bf q} = (\pi,\pi)$ 
charge order, as evidenced by the charge susceptibility $\chi^\mathrm{C}_{\gamma\gamma}({\bf
q})$ plotted in Fig. \ref{Fig:fg2}(b) as a function of temperature.  
Below $1/\beta t_{ps} = 0.2$, the charge correlations
rapidly increase on the $s$ orbital, while there is little change on the $p$ orbitals. 
This observation implies that the charge density on the O sublattice
is uniform, even in the bond disproportionated structure, while a charge
modulation forms on the Bi sites in the insulating state. 
An examination of the real-space charge density, as shown in the inset
of Fig. \ref{Fig:fg2}(b), confirms this. We stress, however, that the charge transfer between the Bi sites is on the order of 0.1 holes/Bi.  

From this analysis, it is clear that the model has a 
bond-disproportionated structure and a small charge modulation on the Bi atoms in the insulating state. This result supports the bond disproportionation scenario proposed for the bismuthates \cite{Foyevtsova_PRB}. 
We now examine how this state evolves upon hole doping. Here, our focus is on the possible formation of lattice polarons, where holes are
bound to local breathing distortions of the oxygen sublattice. 
These objects can be studied by considering the polaron number operator 
$\hat{p}({\bf r})= \hat{x}_{{\bf r},L_s}(\hat{n}_{{\bf r},s}+\hat{n}_{{\bf r},L_s})$,
where $\hat{n}_{{\bf r},L_s} = \sum_\sigma L^\dagger_{{\bf r},s,\sigma}L^{\phantom\dagger}_{{\bf r},s,\sigma}$ 
is the number operator for the $A_{1g}$ combination of the ligand oxygen orbitals  
$L_{{\bf r},s,\sigma} = \frac{1}{2}(p_{{\bf r},x,\sigma} + p_{{\bf r},y,\sigma} - p_{{\bf r},-x,\sigma} - p_{{\bf r},-y,\sigma} )$
~\cite{Supplement} and $\hat{x}_{{\bf r},L_s}=(\hat{X}_{{\bf r},x} + \hat{X}_{{\bf r},y}-\hat{X}_{{\bf r},-x}-\hat{X}_{{\bf r},-y})$. This operator measures the combined presense of  
holes in the $A_{1g}$ molecular orbital surrounding a Bi site and 
a local contraction of those same orbitals, and can be used to trace the 
evolution of polarons with doping. 

With increasing hole concentrations, we observe a MIT at $\beta=10/t_{sp}$. 
Figure \ref{Fig:fg3}(a) plots $\sigma_\mathrm{dc}$ as a function of
filling, where it increases upon hole doping until saturating at $\langle
\hat{n} \rangle \approx 1.4$, indicating metallic behavior. At the same time,
the number of polarons $\frac{1}{N}\sum_{\bf r}\langle \hat{p}({\bf r})\rangle$ 
decreases as additional holes are introduced but remains
nonzero even at the largest dopings [Fig.\ref{Fig:fg3}(b)], 
indicating that the free carriers have polaronic character. 
We also study polaron correlations in real space using  
the staggered polaron correlation function 
$\langle P({\bf r})\rangle =(-1)^{n_x+n_y}\langle \hat{p}({\bf r}) \hat{p}(0) \rangle$, 
which is plotted in Figs.\ref{Fig:fg3}(c)-(f) for selected hole concentrations. At half filling, $\langle
P({\bf r}) \rangle$ is positive for all ${\bf r}$, indicating that the polarons 
are frozen into a long-range two-sublattice order, 
consistent with the patterns inferred from Figs. \ref{Fig:fg1} and \ref{Fig:fg2}. 
With increasing hole concentrations, $\langle P({\bf r}) \rangle$ decreases at the larger distances, 
signalling an overall relaxation of the bond disproportionated on long length scales but the persistence of short-range correlations.  
Such behavior could reflect nanoscale phase separation \cite{Naamneh}; however, 
studies on large clusters are likely needed to clarify this issue. 
Finally, in the high doping region, where the system is metallic 
(e.g. $\langle \hat{n} \rangle > 1.44$), 
the correlations become very short-ranged and extend up to at most one or two lattice constants. 

We also examined the doping evolution of 
the bipolaron number, defined as $\frac{1}{N}\sum_{\bf r} \langle\hat{g}({\bf r})\rangle$, 
where $\hat{g}({\bf r})=\hat{x}_{{\bf r},L_s} 
(\hat{n}_{{\bf r},s,\uparrow}+\hat{n}_{{\bf r},L_s,\uparrow})  
( \hat{n}_{{\bf r},s,\downarrow}+\hat{n}_{{\bf r},L_s,\downarrow})$, and the staggered 
bipolaron correlation function $\langle BP({\bf r}) \rangle = (-1)^{r_x+r_y} 
\langle \hat{g}({\bf r}) \hat{g}(0) \rangle$, as a function of doping.   
When computing the latter quantity, we considered the signal on the 
Bi site by keeping only the terms proportional to 
$\hat{n}_{{\bf r},s,\uparrow}\hat{n}_{{\bf r},s,\downarrow}$. This
simplification is necessary due to the enormous number of terms generated by
the Wick contraction of the product of $\hat{g}({\bf r})$ operators.  
The fact that we see excess charge density on the Bi sites at the center of a
breathing distortion provides some justification for this simplification. 

Figure \ref{Fig:fg3}(b) plots the doping evolution of the bipolaron number operator. As with the polaron number, it is largest near
half-filling and decreases slowly with doping. At large hole concentrations, however, 
it is still finite, suggesting that a significant amount of bipolarons are present in the system. 
The staggered bipolaron correlation function is plotted in Figs.
\ref{Fig:fg3}(g)-(j). At $\langle \hat{n} \rangle =1$, the 
bipolaron correlations are clear and long-ranged on the scale of the cluster. 
This result supports the interpretation that the insulating phase is a static bipolaron lattice. As the hole concentration increases, we find that the bipolaron correlations are suppressed at all length scales, 
while a finite number of bipolarons are present, as indicated in Fig. \ref{Fig:fg3}(b). These results can again be easily understood if the metallic phase is a polaron liquid. 

\begin{figure}[t] \center\includegraphics[width=0.9\columnwidth]{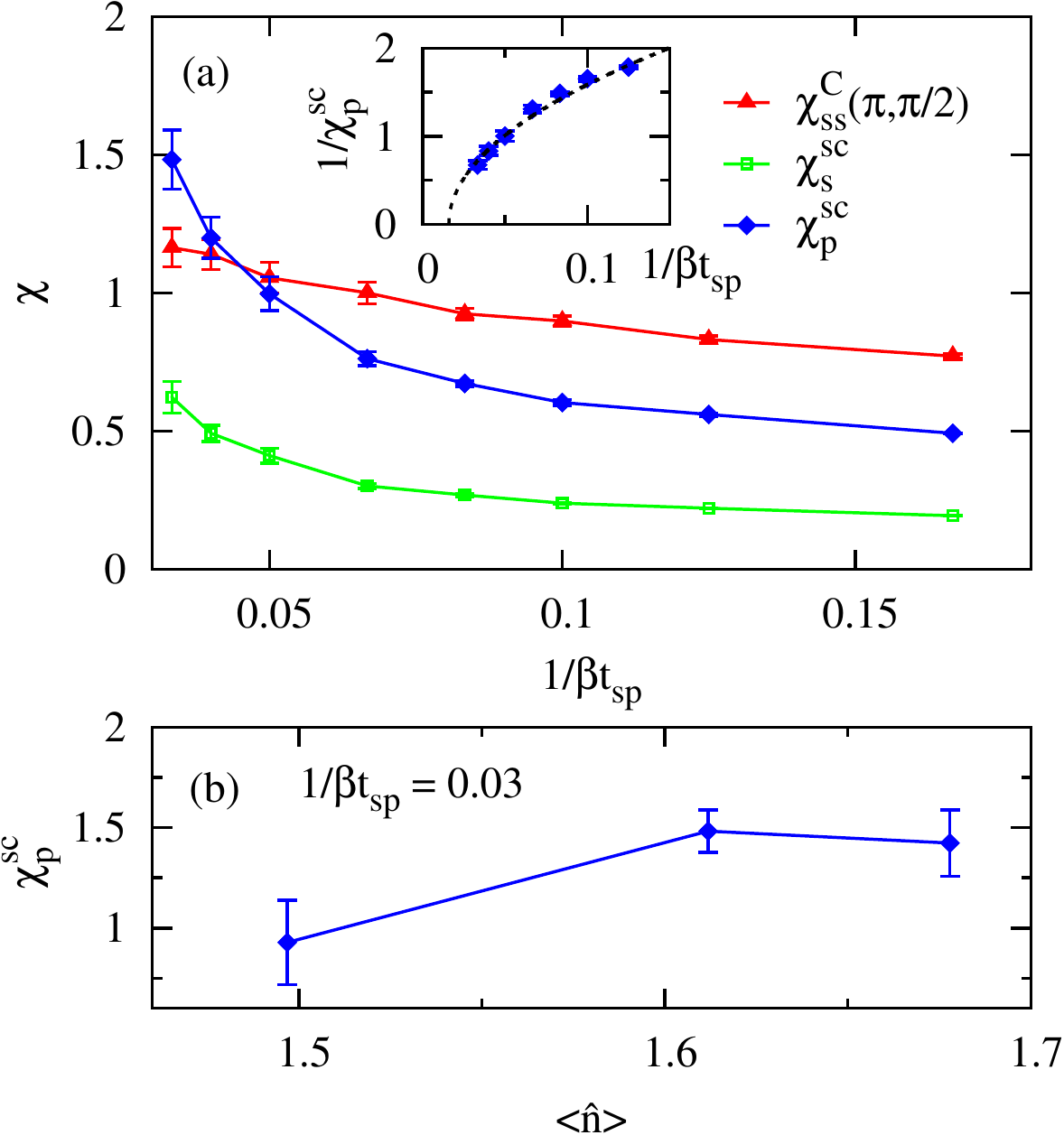}
\caption{\label{Fig:fg4} (a) The charge $\chi_C (\pi,\pi/2)$
and pair-field $\chi_\mathrm{SC}$ susceptibilities as a function of temperature
$1/(\beta t_{sp})$ at $\langle n \rangle = 1.59$. The inset plots
$1/\chi_{sc}^p$ as a function of temperature $1/(\beta t_{sp})$. The black
dashed line is the fitting result. 
(b) The doping dependence of $\chi^p_\mathrm{sc}$ at a 
temperature of $1/(\beta t_{sp}) = 0.03$. 
Error bars smaller than the marker size have
been suppressed for clarity.} \end{figure}

Given the presence of lattice polarons in the metallic phase, we computed the $s$-wave orbital-resolved pair field susceptibility $\chi_{\gamma}^\mathrm{sc}$~\cite{Supplement}.  
Figure \ref{Fig:fg4}(a) plots $\chi_{\gamma}^\mathrm{sc}$ 
as a function of temperature at $\langle \hat{n} \rangle =1.59$, and compares it
against the dominant charge correlations $\chi^{C}_{ss}(\pi,\pi/2)$ at this doping. 
All three susceptibilities increase 
with decreasing temperature, but $\chi^\mathrm{sc}_{p_x} = \chi^\mathrm{sc}_{p_y}\equiv\chi^\mathrm{sc}_p$ 
dominates below $1/\beta t_{sp}\approx 0.04$. 
This observation implies pairing appears predominantly in the oxygen atoms.  
Extrapolating $1/\chi_p^\mathrm{sc}$ to zero (inset), yields 
an estimate for the superconducting 
$\beta_c\approx 63.29/t_{sp}$). This value is artificially high, 
due to the large value of $\Omega$ used in our calculations. Nevertheless, our results provide
evidence that the bipolaronic rich metallic phase has a superconducting ground
state. Fig. \ref{Fig:fg4}(b) plots $\chi^\mathrm{sc}_p$ as a function of doping at $1/\beta t_{sp} = 0.03$, where we find that pairing susceptibility is suppressed in proximity to the insulating phase, suggesting the presence of a superconducting dome induced by competition with the insulating phase.  

{\it Summary} --- We have introduced a quantum Monte Carlo approach for studying bond phonons with 
SSH-type {\eph} couplings in higher dimensions. While our approach has broad applications to many materials, we have used it to study a 2D three-orbital SSH model in the negative charge transfer regime for the first time. We obtained several results consistent with the observed properties of bismuthate high-T$_c$ superconductors. 
At half filling, we find a bond disproportionated state that can be viewed as a lattice of localized bipolarons. Upon hole-doping, this state gives way to a polaron-liquid-like state with short-range correlations, consistent with proposals for nano-scale phase separation or strongly fluctuating lattice polarons in doped BKBO  \cite{IlkaPRB2002, Gallo,PhysRevB.83.174512,PhysRevB.32.6302,Nagata,Naamneh}. 
We also find $s$-wave superconducting tendencies, which primarily form on the oxygen sublattice. 
It would be interesting to contrast our results with those obtained from an effective single band model to fully gauge the importance of the oxygen orbitals. 

{\em Acknowledgments} --- We thank M. Berciu, N. C. Plumb, 
G. A. Sawatzky, and R. T. Scalettar for useful discussions. This work was supported by the Scientific Discovery through Advanced Computing (SciDAC) program funded by the U.S. Department of
Energy, Office of Science, Advanced Scientific Computing Research and Basic Energy Sciences, Division of Materials Sciences and Engineering. 

\bibliography{reference}

\begin{thebibliography}{57}
\expandafter\ifx\csname natexlab\endcsname\relax\def\natexlab#1{#1}\fi
\expandafter\ifx\csname bibnamefont\endcsname\relax
  \def\bibnamefont#1{#1}\fi
\expandafter\ifx\csname bibfnamefont\endcsname\relax
  \def\bibfnamefont#1{#1}\fi
\expandafter\ifx\csname citenamefont\endcsname\relax
  \def\citenamefont#1{#1}\fi
\expandafter\ifx\csname url\endcsname\relax
  \def\url#1{\texttt{#1}}\fi
\expandafter\ifx\csname urlprefix\endcsname\relax\def\urlprefix{URL }\fi
\providecommand{\bibinfo}[2]{#2}
\providecommand{\eprint}[2][]{\url{#2}}

\bibitem[{\citenamefont{Holstein}(1959)}]{Holstein}
\bibinfo{author}{\bibfnamefont{T.}~\bibnamefont{Holstein}},
  \bibinfo{journal}{Annals of Physics} \textbf{\bibinfo{volume}{8}},
  \bibinfo{pages}{325} (\bibinfo{year}{1959}).

\bibitem[{\citenamefont{Fr{\"o}hlich}(1954)}]{Frohlich}
\bibinfo{author}{\bibfnamefont{H.}~\bibnamefont{Fr{\"o}hlich}},
  \bibinfo{journal}{Adv. Phys.} \textbf{\bibinfo{volume}{3}},
  \bibinfo{pages}{325} (\bibinfo{year}{1954}).

\bibitem[{\citenamefont{Su et~al.}(1979)\citenamefont{Su, Schrieffer, and
  Heeger}}]{SuPRL1979}
\bibinfo{author}{\bibfnamefont{W.~P.} \bibnamefont{Su}},
  \bibinfo{author}{\bibfnamefont{J.~R.} \bibnamefont{Schrieffer}},
  \bibnamefont{and} \bibinfo{author}{\bibfnamefont{A.~J.}
  \bibnamefont{Heeger}}, \bibinfo{journal}{Phys. Rev. Lett.}
  \textbf{\bibinfo{volume}{42}}, \bibinfo{pages}{1698} (\bibinfo{year}{1979}),
  \urlprefix\url{https://link.aps.org/doi/10.1103/PhysRevLett.42.1698}.

\bibitem[{\citenamefont{Sangiovanni et~al.}(2006)\citenamefont{Sangiovanni,
  Gunnarsson, Koch, Castellani, and Capone}}]{Sangiovanni}
\bibinfo{author}{\bibfnamefont{G.}~\bibnamefont{Sangiovanni}},
  \bibinfo{author}{\bibfnamefont{O.}~\bibnamefont{Gunnarsson}},
  \bibinfo{author}{\bibfnamefont{E.}~\bibnamefont{Koch}},
  \bibinfo{author}{\bibfnamefont{C.}~\bibnamefont{Castellani}},
  \bibnamefont{and} \bibinfo{author}{\bibfnamefont{M.}~\bibnamefont{Capone}},
  \bibinfo{journal}{Phys. Rev. Lett.} \textbf{\bibinfo{volume}{97}},
  \bibinfo{pages}{046404} (\bibinfo{year}{2006}),
  \urlprefix\url{https://link.aps.org/doi/10.1103/PhysRevLett.97.046404}.

\bibitem[{\citenamefont{Macridin et~al.}(2006)\citenamefont{Macridin, Moritz,
  Jarrell, and Maier}}]{Macridin}
\bibinfo{author}{\bibfnamefont{A.}~\bibnamefont{Macridin}},
  \bibinfo{author}{\bibfnamefont{B.}~\bibnamefont{Moritz}},
  \bibinfo{author}{\bibfnamefont{M.}~\bibnamefont{Jarrell}}, \bibnamefont{and}
  \bibinfo{author}{\bibfnamefont{T.}~\bibnamefont{Maier}},
  \bibinfo{journal}{Phys. Rev. Lett.} \textbf{\bibinfo{volume}{97}},
  \bibinfo{pages}{056402} (\bibinfo{year}{2006}),
  \urlprefix\url{https://link.aps.org/doi/10.1103/PhysRevLett.97.056402}.

\bibitem[{\citenamefont{Bauer}(2010)}]{Bauer2010}
\bibinfo{author}{\bibfnamefont{J.}~\bibnamefont{Bauer}}, \bibinfo{journal}{EPL}
  \textbf{\bibinfo{volume}{90}}, \bibinfo{pages}{27002} (\bibinfo{year}{2010}),
  \urlprefix\url{http://stacks.iop.org/0295-5075/90/i=2/a=27002}.

\bibitem[{\citenamefont{Bauer and Hewson}(2010)}]{BauerPRB}
\bibinfo{author}{\bibfnamefont{J.}~\bibnamefont{Bauer}} \bibnamefont{and}
  \bibinfo{author}{\bibfnamefont{A.~C.} \bibnamefont{Hewson}},
  \bibinfo{journal}{Phys. Rev. B} \textbf{\bibinfo{volume}{81}},
  \bibinfo{pages}{235113} (\bibinfo{year}{2010}),
  \urlprefix\url{https://link.aps.org/doi/10.1103/PhysRevB.81.235113}.

\bibitem[{\citenamefont{Nowadnick et~al.}(2012)\citenamefont{Nowadnick,
  Johnston, Moritz, Scalettar, and Devereaux}}]{Nowadnick}
\bibinfo{author}{\bibfnamefont{E.~A.} \bibnamefont{Nowadnick}},
  \bibinfo{author}{\bibfnamefont{S.}~\bibnamefont{Johnston}},
  \bibinfo{author}{\bibfnamefont{B.}~\bibnamefont{Moritz}},
  \bibinfo{author}{\bibfnamefont{R.~T.} \bibnamefont{Scalettar}},
  \bibnamefont{and} \bibinfo{author}{\bibfnamefont{T.~P.}
  \bibnamefont{Devereaux}}, \bibinfo{journal}{Phys. Rev. Lett.}
  \textbf{\bibinfo{volume}{109}}, \bibinfo{pages}{246404}
  (\bibinfo{year}{2012}),
  \urlprefix\url{https://link.aps.org/doi/10.1103/PhysRevLett.109.246404}.

\bibitem[{\citenamefont{Murakami et~al.}(2013)\citenamefont{Murakami, Werner,
  Tsuji, and Aoki}}]{MurakamiPRB}
\bibinfo{author}{\bibfnamefont{Y.}~\bibnamefont{Murakami}},
  \bibinfo{author}{\bibfnamefont{P.}~\bibnamefont{Werner}},
  \bibinfo{author}{\bibfnamefont{N.}~\bibnamefont{Tsuji}}, \bibnamefont{and}
  \bibinfo{author}{\bibfnamefont{H.}~\bibnamefont{Aoki}},
  \bibinfo{journal}{Phys. Rev. B} \textbf{\bibinfo{volume}{88}},
  \bibinfo{pages}{125126} (\bibinfo{year}{2013}),
  \urlprefix\url{https://link.aps.org/doi/10.1103/PhysRevB.88.125126}.

\bibitem[{\citenamefont{Li et~al.}(2017)\citenamefont{Li, Khatami, and
  Johnston}}]{ShaozhiPRB2017}
\bibinfo{author}{\bibfnamefont{S.}~\bibnamefont{Li}},
  \bibinfo{author}{\bibfnamefont{E.}~\bibnamefont{Khatami}}, \bibnamefont{and}
  \bibinfo{author}{\bibfnamefont{S.}~\bibnamefont{Johnston}},
  \bibinfo{journal}{Phys. Rev. B} \textbf{\bibinfo{volume}{95}},
  \bibinfo{pages}{121112} (\bibinfo{year}{2017}),
  \urlprefix\url{https://link.aps.org/doi/10.1103/PhysRevB.95.121112}.

\bibitem[{\citenamefont{Berciu}(2006)}]{BerciuPRL}
\bibinfo{author}{\bibfnamefont{M.}~\bibnamefont{Berciu}},
  \bibinfo{journal}{Phys. Rev. Lett.} \textbf{\bibinfo{volume}{97}},
  \bibinfo{pages}{036402} (\bibinfo{year}{2006}),
  \urlprefix\url{https://link.aps.org/doi/10.1103/PhysRevLett.97.036402}.

\bibitem[{\citenamefont{Noack et~al.}(1991)\citenamefont{Noack, Scalapino, and
  Scalettar}}]{NoackPRL}
\bibinfo{author}{\bibfnamefont{R.~M.} \bibnamefont{Noack}},
  \bibinfo{author}{\bibfnamefont{D.~J.} \bibnamefont{Scalapino}},
  \bibnamefont{and} \bibinfo{author}{\bibfnamefont{R.~T.}
  \bibnamefont{Scalettar}}, \bibinfo{journal}{Phys. Rev. Lett.}
  \textbf{\bibinfo{volume}{66}}, \bibinfo{pages}{778} (\bibinfo{year}{1991}),
  \urlprefix\url{https://link.aps.org/doi/10.1103/PhysRevLett.66.778}.

\bibitem[{\citenamefont{Costa et~al.}(2018)\citenamefont{Costa, Blommel, Chiu,
  Batrouni, and Scalettar}}]{Costa2018}
\bibinfo{author}{\bibfnamefont{N.~C.} \bibnamefont{Costa}},
  \bibinfo{author}{\bibfnamefont{T.}~\bibnamefont{Blommel}},
  \bibinfo{author}{\bibfnamefont{W.-T.} \bibnamefont{Chiu}},
  \bibinfo{author}{\bibfnamefont{G.}~\bibnamefont{Batrouni}}, \bibnamefont{and}
  \bibinfo{author}{\bibfnamefont{R.~T.} \bibnamefont{Scalettar}},
  \bibinfo{journal}{Phys. Rev. Lett.} \textbf{\bibinfo{volume}{120}},
  \bibinfo{pages}{187003} (\bibinfo{year}{2018}),
  \urlprefix\url{https://link.aps.org/doi/10.1103/PhysRevLett.120.187003}.

\bibitem[{\citenamefont{Bon\ifmmode~\check{c}\else \v{c}\fi{}a
  et~al.}(1999)\citenamefont{Bon\ifmmode~\check{c}\else \v{c}\fi{}a, Trugman,
  and Batisti\ifmmode~\acute{c}\else \'{c}\fi{}}}]{Bonca}
\bibinfo{author}{\bibfnamefont{J.}~\bibnamefont{Bon\ifmmode~\check{c}\else
  \v{c}\fi{}a}}, \bibinfo{author}{\bibfnamefont{S.~A.} \bibnamefont{Trugman}},
  \bibnamefont{and}
  \bibinfo{author}{\bibfnamefont{I.}~\bibnamefont{Batisti\ifmmode~\acute{c}\else
  \'{c}\fi{}}}, \bibinfo{journal}{Phys. Rev. B} \textbf{\bibinfo{volume}{60}},
  \bibinfo{pages}{1633} (\bibinfo{year}{1999}),
  \urlprefix\url{https://link.aps.org/doi/10.1103/PhysRevB.60.1633}.

\bibitem[{\citenamefont{Ku et~al.}(2002)\citenamefont{Ku, Trugman, and
  Bon\ifmmode~\check{c}\else \v{c}\fi{}a}}]{KuPRB2002}
\bibinfo{author}{\bibfnamefont{L.-C.} \bibnamefont{Ku}},
  \bibinfo{author}{\bibfnamefont{S.~A.} \bibnamefont{Trugman}},
  \bibnamefont{and}
  \bibinfo{author}{\bibfnamefont{J.}~\bibnamefont{Bon\ifmmode~\check{c}\else
  \v{c}\fi{}a}}, \bibinfo{journal}{Phys. Rev. B} \textbf{\bibinfo{volume}{65}},
  \bibinfo{pages}{174306} (\bibinfo{year}{2002}),
  \urlprefix\url{https://link.aps.org/doi/10.1103/PhysRevB.65.174306}.

\bibitem[{\citenamefont{Hague et~al.}(2006)\citenamefont{Hague, Kornilovitch,
  Alexandrov, and Samson}}]{HaguePRB2006}
\bibinfo{author}{\bibfnamefont{J.~P.} \bibnamefont{Hague}},
  \bibinfo{author}{\bibfnamefont{P.~E.} \bibnamefont{Kornilovitch}},
  \bibinfo{author}{\bibfnamefont{A.~S.} \bibnamefont{Alexandrov}},
  \bibnamefont{and} \bibinfo{author}{\bibfnamefont{J.~H.}
  \bibnamefont{Samson}}, \bibinfo{journal}{Phys. Rev. B}
  \textbf{\bibinfo{volume}{73}}, \bibinfo{pages}{054303}
  (\bibinfo{year}{2006}),
  \urlprefix\url{https://link.aps.org/doi/10.1103/PhysRevB.73.054303}.

\bibitem[{\citenamefont{Romero et~al.}(1999)\citenamefont{Romero, Brown, and
  Lindenberg}}]{RomeroPRB}
\bibinfo{author}{\bibfnamefont{A.~H.} \bibnamefont{Romero}},
  \bibinfo{author}{\bibfnamefont{D.~W.} \bibnamefont{Brown}}, \bibnamefont{and}
  \bibinfo{author}{\bibfnamefont{K.}~\bibnamefont{Lindenberg}},
  \bibinfo{journal}{Phys. Rev. B} \textbf{\bibinfo{volume}{59}},
  \bibinfo{pages}{13728} (\bibinfo{year}{1999}),
  \urlprefix\url{https://link.aps.org/doi/10.1103/PhysRevB.59.13728}.

\bibitem[{\citenamefont{Li et~al.}(2015)\citenamefont{Li, Nowadnick, and
  Johnston}}]{shaozhiPRB2015}
\bibinfo{author}{\bibfnamefont{S.}~\bibnamefont{Li}},
  \bibinfo{author}{\bibfnamefont{E.~A.} \bibnamefont{Nowadnick}},
  \bibnamefont{and} \bibinfo{author}{\bibfnamefont{S.}~\bibnamefont{Johnston}},
  \bibinfo{journal}{Phys. Rev. B} \textbf{\bibinfo{volume}{92}},
  \bibinfo{pages}{064301} (\bibinfo{year}{2015}),
  \urlprefix\url{https://link.aps.org/doi/10.1103/PhysRevB.92.064301}.

\bibitem[{\citenamefont{Chakraverty et~al.}(1998)\citenamefont{Chakraverty,
  Ranninger, and Feinberg}}]{Chakraverty}
\bibinfo{author}{\bibfnamefont{B.~K.} \bibnamefont{Chakraverty}},
  \bibinfo{author}{\bibfnamefont{J.}~\bibnamefont{Ranninger}},
  \bibnamefont{and} \bibinfo{author}{\bibfnamefont{D.}~\bibnamefont{Feinberg}},
  \bibinfo{journal}{Phys. Rev. Lett.} \textbf{\bibinfo{volume}{81}},
  \bibinfo{pages}{433} (\bibinfo{year}{1998}),
  \urlprefix\url{https://link.aps.org/doi/10.1103/PhysRevLett.81.433}.

\bibitem[{\citenamefont{Prokof'ev and Svistunov}(1998)}]{Diagrams}
\bibinfo{author}{\bibfnamefont{N.~V.} \bibnamefont{Prokof'ev}}
  \bibnamefont{and} \bibinfo{author}{\bibfnamefont{B.~V.}
  \bibnamefont{Svistunov}}, \bibinfo{journal}{Phys. Rev. Lett.}
  \textbf{\bibinfo{volume}{81}}, \bibinfo{pages}{2514} (\bibinfo{year}{1998}),
  \urlprefix\url{https://link.aps.org/doi/10.1103/PhysRevLett.81.2514}.

\bibitem[{\citenamefont{Weber and Hohenadler}(2018)}]{HohenadlerPRB2018}
\bibinfo{author}{\bibfnamefont{M.}~\bibnamefont{Weber}} \bibnamefont{and}
  \bibinfo{author}{\bibfnamefont{M.}~\bibnamefont{Hohenadler}},
  \bibinfo{journal}{Phys. Rev. B} \textbf{\bibinfo{volume}{98}},
  \bibinfo{pages}{085405} (\bibinfo{year}{2018}),
  \urlprefix\url{https://link.aps.org/doi/10.1103/PhysRevB.98.085405}.

\bibitem[{\citenamefont{Dee et~al.}(2019)\citenamefont{Dee, Nakatsukasa, Wang,
  and Johnston}}]{DeePRB2019}
\bibinfo{author}{\bibfnamefont{P.~M.} \bibnamefont{Dee}},
  \bibinfo{author}{\bibfnamefont{K.}~\bibnamefont{Nakatsukasa}},
  \bibinfo{author}{\bibfnamefont{Y.}~\bibnamefont{Wang}}, \bibnamefont{and}
  \bibinfo{author}{\bibfnamefont{S.}~\bibnamefont{Johnston}},
  \bibinfo{journal}{arXiv:1811.03676}  (\bibinfo{year}{2019}).

\bibitem[{\citenamefont{Li et~al.}(2013)\citenamefont{Li, Dong, Yi, and
  Xie}}]{Shaozhi2013}
\bibinfo{author}{\bibfnamefont{S.}~\bibnamefont{Li}},
  \bibinfo{author}{\bibfnamefont{X.}~\bibnamefont{Dong}},
  \bibinfo{author}{\bibfnamefont{D.}~\bibnamefont{Yi}}, \bibnamefont{and}
  \bibinfo{author}{\bibfnamefont{S.}~\bibnamefont{Xie}},
  \bibinfo{journal}{Organic Electronics} \textbf{\bibinfo{volume}{14}},
  \bibinfo{pages}{2216 } (\bibinfo{year}{2013}), ISSN
  \bibinfo{issn}{1566-1199},
  \urlprefix\url{http://www.sciencedirect.com/science/article/pii/S1566119913002565}.

\bibitem[{\citenamefont{Marchand et~al.}(2010)\citenamefont{Marchand,
  De~Filippis, Cataudella, Berciu, Nagaosa, Prokof'ev, Mishchenko, and
  Stamp}}]{Marchand}
\bibinfo{author}{\bibfnamefont{D.~J.~J.} \bibnamefont{Marchand}},
  \bibinfo{author}{\bibfnamefont{G.}~\bibnamefont{De~Filippis}},
  \bibinfo{author}{\bibfnamefont{V.}~\bibnamefont{Cataudella}},
  \bibinfo{author}{\bibfnamefont{M.}~\bibnamefont{Berciu}},
  \bibinfo{author}{\bibfnamefont{N.}~\bibnamefont{Nagaosa}},
  \bibinfo{author}{\bibfnamefont{N.~V.} \bibnamefont{Prokof'ev}},
  \bibinfo{author}{\bibfnamefont{A.~S.} \bibnamefont{Mishchenko}},
  \bibnamefont{and} \bibinfo{author}{\bibfnamefont{P.~C.~E.}
  \bibnamefont{Stamp}}, \bibinfo{journal}{Phys. Rev. Lett.}
  \textbf{\bibinfo{volume}{105}}, \bibinfo{pages}{266605}
  (\bibinfo{year}{2010}),
  \urlprefix\url{https://link.aps.org/doi/10.1103/PhysRevLett.105.266605}.

\bibitem[{\citenamefont{Sous et~al.}(2018)\citenamefont{Sous, Chakaraborty,
  Krems, and Berciu}}]{Sous}
\bibinfo{author}{\bibfnamefont{J.}~\bibnamefont{Sous}},
  \bibinfo{author}{\bibfnamefont{M.}~\bibnamefont{Chakaraborty}},
  \bibinfo{author}{\bibfnamefont{R.~V.} \bibnamefont{Krems}}, \bibnamefont{and}
  \bibinfo{author}{\bibfnamefont{M.}~\bibnamefont{Berciu}},
  \bibinfo{journal}{arXiv:1805.06109}  (\bibinfo{year}{2018}).

\bibitem[{\citenamefont{Sengupta et~al.}(2003)\citenamefont{Sengupta, Sandvik,
  and Campbell}}]{SenguptaPRB}
\bibinfo{author}{\bibfnamefont{P.}~\bibnamefont{Sengupta}},
  \bibinfo{author}{\bibfnamefont{A.~W.} \bibnamefont{Sandvik}},
  \bibnamefont{and} \bibinfo{author}{\bibfnamefont{D.~K.}
  \bibnamefont{Campbell}}, \bibinfo{journal}{Phys. Rev. B}
  \textbf{\bibinfo{volume}{67}}, \bibinfo{pages}{245103}
  (\bibinfo{year}{2003}),
  \urlprefix\url{https://link.aps.org/doi/10.1103/PhysRevB.67.245103}.

\bibitem[{\citenamefont{Hohenadler}(2016)}]{Hohenadler2016}
\bibinfo{author}{\bibfnamefont{M.}~\bibnamefont{Hohenadler}},
  \bibinfo{journal}{Phys. Rev. Lett.} \textbf{\bibinfo{volume}{117}},
  \bibinfo{pages}{206404} (\bibinfo{year}{2016}),
  \urlprefix\url{https://link.aps.org/doi/10.1103/PhysRevLett.117.206404}.

\bibitem[{\citenamefont{Tang and Hirsch}(1988)}]{TangPRB1988}
\bibinfo{author}{\bibfnamefont{S.}~\bibnamefont{Tang}} \bibnamefont{and}
  \bibinfo{author}{\bibfnamefont{J.~E.} \bibnamefont{Hirsch}},
  \bibinfo{journal}{Phys. Rev. B} \textbf{\bibinfo{volume}{37}},
  \bibinfo{pages}{9546} (\bibinfo{year}{1988}),
  \urlprefix\url{https://link.aps.org/doi/10.1103/PhysRevB.37.9546}.

\bibitem[{\citenamefont{Clay and Mazumdar}(2018)}]{Clay2018}
\bibinfo{author}{\bibfnamefont{R.~T.} \bibnamefont{Clay}} \bibnamefont{and}
  \bibinfo{author}{\bibfnamefont{S.}~\bibnamefont{Mazumdar}}
  (\bibinfo{year}{2018}), \eprint{arXiv:1802.01551}.

\bibitem[{\citenamefont{Medarde}(1996)}]{Medarde}
\bibinfo{author}{\bibfnamefont{M.~L.} \bibnamefont{Medarde}},
  \bibinfo{journal}{Journal of Physics: Condensed Matter}
  \textbf{\bibinfo{volume}{9}}, \bibinfo{pages}{1679} (\bibinfo{year}{1996}).

\bibitem[{\citenamefont{Shamblin et~al.}(2018)\citenamefont{Shamblin, Heres,
  Zhou, Sangoro, Lang, Neuefeind, Alonso, and Johnston}}]{Shamblin2018}
\bibinfo{author}{\bibfnamefont{J.}~\bibnamefont{Shamblin}},
  \bibinfo{author}{\bibfnamefont{M.}~\bibnamefont{Heres}},
  \bibinfo{author}{\bibfnamefont{H.}~\bibnamefont{Zhou}},
  \bibinfo{author}{\bibfnamefont{J.}~\bibnamefont{Sangoro}},
  \bibinfo{author}{\bibfnamefont{M.}~\bibnamefont{Lang}},
  \bibinfo{author}{\bibfnamefont{J.}~\bibnamefont{Neuefeind}},
  \bibinfo{author}{\bibfnamefont{J.~A.} \bibnamefont{Alonso}},
  \bibnamefont{and} \bibinfo{author}{\bibfnamefont{S.}~\bibnamefont{Johnston}},
  \bibinfo{journal}{Nat. Commun.} \textbf{\bibinfo{volume}{9}},
  \bibinfo{pages}{86} (\bibinfo{year}{2018}),
  \urlprefix\url{https://doi.org/10.1038/s41467-017-02561-6}.

\bibitem[{\citenamefont{Johnston et~al.}(2014)\citenamefont{Johnston,
  Mukherjee, Elfimov, Berciu, and Sawatzky}}]{JohnstonPRL}
\bibinfo{author}{\bibfnamefont{S.}~\bibnamefont{Johnston}},
  \bibinfo{author}{\bibfnamefont{A.}~\bibnamefont{Mukherjee}},
  \bibinfo{author}{\bibfnamefont{I.}~\bibnamefont{Elfimov}},
  \bibinfo{author}{\bibfnamefont{M.}~\bibnamefont{Berciu}}, \bibnamefont{and}
  \bibinfo{author}{\bibfnamefont{G.~A.} \bibnamefont{Sawatzky}},
  \bibinfo{journal}{Phys. Rev. Lett.} \textbf{\bibinfo{volume}{112}},
  \bibinfo{pages}{106404} (\bibinfo{year}{2014}),
  \urlprefix\url{https://link.aps.org/doi/10.1103/PhysRevLett.112.106404}.

\bibitem[{\citenamefont{Lanzara et~al.}(2001)\citenamefont{Lanzara, Bogdanov,
  Zhou, Kellar, Feng, Lu, Yoshida, Eisaki, Fujimori, Kishio
  et~al.}}]{Lanzara_Nature}
\bibinfo{author}{\bibfnamefont{A.}~\bibnamefont{Lanzara}},
  \bibinfo{author}{\bibfnamefont{P.~V.} \bibnamefont{Bogdanov}},
  \bibinfo{author}{\bibfnamefont{X.~J.} \bibnamefont{Zhou}},
  \bibinfo{author}{\bibfnamefont{S.~A.} \bibnamefont{Kellar}},
  \bibinfo{author}{\bibfnamefont{D.~L.} \bibnamefont{Feng}},
  \bibinfo{author}{\bibfnamefont{E.~D.} \bibnamefont{Lu}},
  \bibinfo{author}{\bibfnamefont{T.}~\bibnamefont{Yoshida}},
  \bibinfo{author}{\bibfnamefont{H.}~\bibnamefont{Eisaki}},
  \bibinfo{author}{\bibfnamefont{A.}~\bibnamefont{Fujimori}},
  \bibinfo{author}{\bibfnamefont{K.}~\bibnamefont{Kishio}},
  \bibnamefont{et~al.}, \bibinfo{journal}{Nature}
  \textbf{\bibinfo{volume}{412}}, \bibinfo{pages}{510} (\bibinfo{year}{2001}).

\bibitem[{\citenamefont{Weber}(1987)}]{Weber_PRL}
\bibinfo{author}{\bibfnamefont{W.}~\bibnamefont{Weber}},
  \bibinfo{journal}{Phys. Rev. Lett.} \textbf{\bibinfo{volume}{58}},
  \bibinfo{pages}{1371} (\bibinfo{year}{1987}),
  \urlprefix\url{https://link.aps.org/doi/10.1103/PhysRevLett.58.1371}.

\bibitem[{\citenamefont{Khazraie et~al.}(2018)\citenamefont{Khazraie,
  Foyevtsova, Elfimov, and Sawatzky}}]{KhazraiePRB}
\bibinfo{author}{\bibfnamefont{A.}~\bibnamefont{Khazraie}},
  \bibinfo{author}{\bibfnamefont{K.}~\bibnamefont{Foyevtsova}},
  \bibinfo{author}{\bibfnamefont{I.}~\bibnamefont{Elfimov}}, \bibnamefont{and}
  \bibinfo{author}{\bibfnamefont{G.~A.} \bibnamefont{Sawatzky}},
  \bibinfo{journal}{Phys. Rev. B} \textbf{\bibinfo{volume}{97}},
  \bibinfo{pages}{075103} (\bibinfo{year}{2018}),
  \urlprefix\url{https://link.aps.org/doi/10.1103/PhysRevB.97.075103}.

\bibitem[{\citenamefont{M\"{o}ller et~al.}(2017)\citenamefont{M\"{o}ller,
  Sawatzky, Franz, and Berciu}}]{Moeller2017}
\bibinfo{author}{\bibfnamefont{M.~M.} \bibnamefont{M\"{o}ller}},
  \bibinfo{author}{\bibfnamefont{G.~A.} \bibnamefont{Sawatzky}},
  \bibinfo{author}{\bibfnamefont{M.}~\bibnamefont{Franz}}, \bibnamefont{and}
  \bibinfo{author}{\bibfnamefont{M.}~\bibnamefont{Berciu}},
  \bibinfo{journal}{Nat. Commun.} \textbf{\bibinfo{volume}{8}},
  \bibinfo{pages}{2267} (\bibinfo{year}{2017}),
  \urlprefix\url{https://doi.org/10.1038/s41467-017-02442-y}.

\bibitem[{\citenamefont{Qi and Zhang}(2011)}]{QiRMP2011}
\bibinfo{author}{\bibfnamefont{X.-L.} \bibnamefont{Qi}} \bibnamefont{and}
  \bibinfo{author}{\bibfnamefont{S.-C.} \bibnamefont{Zhang}},
  \bibinfo{journal}{Rev. Mod. Phys.} \textbf{\bibinfo{volume}{83}},
  \bibinfo{pages}{1057} (\bibinfo{year}{2011}),
  \urlprefix\url{https://link.aps.org/doi/10.1103/RevModPhys.83.1057}.

\bibitem[{\citenamefont{Schnyder et~al.}(2008)\citenamefont{Schnyder, Ryu,
  Furusaki, and Ludwig}}]{SchnyderPRB2008}
\bibinfo{author}{\bibfnamefont{A.~P.} \bibnamefont{Schnyder}},
  \bibinfo{author}{\bibfnamefont{S.}~\bibnamefont{Ryu}},
  \bibinfo{author}{\bibfnamefont{A.}~\bibnamefont{Furusaki}}, \bibnamefont{and}
  \bibinfo{author}{\bibfnamefont{A.~W.~W.} \bibnamefont{Ludwig}},
  \bibinfo{journal}{Phys. Rev. B} \textbf{\bibinfo{volume}{78}},
  \bibinfo{pages}{195125} (\bibinfo{year}{2008}),
  \urlprefix\url{https://link.aps.org/doi/10.1103/PhysRevB.78.195125}.

\bibitem[{\citenamefont{Mizokawa et~al.}(1991)\citenamefont{Mizokawa, Namatame,
  Fujimori, Akeyama, Kondoh, Kuroda, and Kosugi}}]{Mizokawa}
\bibinfo{author}{\bibfnamefont{T.}~\bibnamefont{Mizokawa}},
  \bibinfo{author}{\bibfnamefont{H.}~\bibnamefont{Namatame}},
  \bibinfo{author}{\bibfnamefont{A.}~\bibnamefont{Fujimori}},
  \bibinfo{author}{\bibfnamefont{K.}~\bibnamefont{Akeyama}},
  \bibinfo{author}{\bibfnamefont{H.}~\bibnamefont{Kondoh}},
  \bibinfo{author}{\bibfnamefont{H.}~\bibnamefont{Kuroda}}, \bibnamefont{and}
  \bibinfo{author}{\bibfnamefont{N.}~\bibnamefont{Kosugi}},
  \bibinfo{journal}{Phys. Rev. Lett.} \textbf{\bibinfo{volume}{67}},
  \bibinfo{pages}{1638} (\bibinfo{year}{1991}),
  \urlprefix\url{https://link.aps.org/doi/10.1103/PhysRevLett.67.1638}.

\bibitem[{\citenamefont{Zaanen et~al.}(1985)\citenamefont{Zaanen, Sawatzky, and
  Allen}}]{ZSA}
\bibinfo{author}{\bibfnamefont{J.}~\bibnamefont{Zaanen}},
  \bibinfo{author}{\bibfnamefont{G.~A.} \bibnamefont{Sawatzky}},
  \bibnamefont{and} \bibinfo{author}{\bibfnamefont{J.~W.} \bibnamefont{Allen}},
  \bibinfo{journal}{Phys. Rev. Lett.} \textbf{\bibinfo{volume}{55}},
  \bibinfo{pages}{418} (\bibinfo{year}{1985}),
  \urlprefix\url{https://link.aps.org/doi/10.1103/PhysRevLett.55.418}.

\bibitem[{\citenamefont{Foyevtsova et~al.}(2015)\citenamefont{Foyevtsova,
  Khazraie, Elfimov, and Sawatzky}}]{Foyevtsova_PRB}
\bibinfo{author}{\bibfnamefont{K.}~\bibnamefont{Foyevtsova}},
  \bibinfo{author}{\bibfnamefont{A.}~\bibnamefont{Khazraie}},
  \bibinfo{author}{\bibfnamefont{I.}~\bibnamefont{Elfimov}}, \bibnamefont{and}
  \bibinfo{author}{\bibfnamefont{G.~A.} \bibnamefont{Sawatzky}},
  \bibinfo{journal}{Phys. Rev. B} \textbf{\bibinfo{volume}{91}},
  \bibinfo{pages}{121114} (\bibinfo{year}{2015}),
  \urlprefix\url{https://link.aps.org/doi/10.1103/PhysRevB.91.121114}.

\bibitem[{\citenamefont{Plumb et~al.}(2016)\citenamefont{Plumb, Gawryluk, Wang,
  Risti\ifmmode~\acute{c}\else \'{c}\fi{}, Park, Lv, Wang, Matt, Xu, Shang
  et~al.}}]{Plumb2016}
\bibinfo{author}{\bibfnamefont{N.~C.} \bibnamefont{Plumb}},
  \bibinfo{author}{\bibfnamefont{D.~J.} \bibnamefont{Gawryluk}},
  \bibinfo{author}{\bibfnamefont{Y.}~\bibnamefont{Wang}},
  \bibinfo{author}{\bibfnamefont{Z.}~\bibnamefont{Risti\ifmmode~\acute{c}\else
  \'{c}\fi{}}}, \bibinfo{author}{\bibfnamefont{J.}~\bibnamefont{Park}},
  \bibinfo{author}{\bibfnamefont{B.~Q.} \bibnamefont{Lv}},
  \bibinfo{author}{\bibfnamefont{Z.}~\bibnamefont{Wang}},
  \bibinfo{author}{\bibfnamefont{C.~E.} \bibnamefont{Matt}},
  \bibinfo{author}{\bibfnamefont{N.}~\bibnamefont{Xu}},
  \bibinfo{author}{\bibfnamefont{T.}~\bibnamefont{Shang}},
  \bibnamefont{et~al.}, \bibinfo{journal}{Phys. Rev. Lett.}
  \textbf{\bibinfo{volume}{117}}, \bibinfo{pages}{037002}
  (\bibinfo{year}{2016}),
  \urlprefix\url{https://link.aps.org/doi/10.1103/PhysRevLett.117.037002}.

\bibitem[{\citenamefont{Yin et~al.}(2013)\citenamefont{Yin, Kutepov, and
  Kotliar}}]{PhysRevX.3.021011}
\bibinfo{author}{\bibfnamefont{Z.~P.} \bibnamefont{Yin}},
  \bibinfo{author}{\bibfnamefont{A.}~\bibnamefont{Kutepov}}, \bibnamefont{and}
  \bibinfo{author}{\bibfnamefont{G.}~\bibnamefont{Kotliar}},
  \bibinfo{journal}{Phys. Rev. X} \textbf{\bibinfo{volume}{3}},
  \bibinfo{pages}{021011} (\bibinfo{year}{2013}),
  \urlprefix\url{https://link.aps.org/doi/10.1103/PhysRevX.3.021011}.

\bibitem[{\citenamefont{Park et~al.}(2012)\citenamefont{Park, Millis, and
  Marianetti}}]{Park2012}
\bibinfo{author}{\bibfnamefont{H.}~\bibnamefont{Park}},
  \bibinfo{author}{\bibfnamefont{A.~J.} \bibnamefont{Millis}},
  \bibnamefont{and} \bibinfo{author}{\bibfnamefont{C.~A.}
  \bibnamefont{Marianetti}}, \bibinfo{journal}{Phys. Rev. Lett.}
  \textbf{\bibinfo{volume}{109}}, \bibinfo{pages}{156402}
  (\bibinfo{year}{2012}),
  \urlprefix\url{https://link.aps.org/doi/10.1103/PhysRevLett.109.156402}.

\bibitem[{\citenamefont{Bisogni et~al.}(2016)\citenamefont{Bisogni, Catalano,
  Green, Gibert, Scherwitzl, Huang, Strocov, Zubko, Balandeh, Triscone
  et~al.}}]{Bisogni2016}
\bibinfo{author}{\bibfnamefont{V.}~\bibnamefont{Bisogni}},
  \bibinfo{author}{\bibfnamefont{S.}~\bibnamefont{Catalano}},
  \bibinfo{author}{\bibfnamefont{R.~J.} \bibnamefont{Green}},
  \bibinfo{author}{\bibfnamefont{M.}~\bibnamefont{Gibert}},
  \bibinfo{author}{\bibfnamefont{R.}~\bibnamefont{Scherwitzl}},
  \bibinfo{author}{\bibfnamefont{Y.}~\bibnamefont{Huang}},
  \bibinfo{author}{\bibfnamefont{V.~N.} \bibnamefont{Strocov}},
  \bibinfo{author}{\bibfnamefont{P.}~\bibnamefont{Zubko}},
  \bibinfo{author}{\bibfnamefont{S.}~\bibnamefont{Balandeh}},
  \bibinfo{author}{\bibfnamefont{J.-M.} \bibnamefont{Triscone}},
  \bibnamefont{et~al.}, \bibinfo{journal}{Nature Communications}
  \textbf{\bibinfo{volume}{7}}, \bibinfo{pages}{13017} (\bibinfo{year}{2016}).

\bibitem[{\citenamefont{Sleight}(2015)}]{Sleight}
\bibinfo{author}{\bibfnamefont{A.~W.} \bibnamefont{Sleight}},
  \bibinfo{journal}{Physica C: Superconductivity and its Applications}
  \textbf{\bibinfo{volume}{214}}, \bibinfo{pages}{152} (\bibinfo{year}{2015}).

\bibitem[{Sup()}]{Supplement}
\bibinfo{journal}{See supplementary materials at $\dots$.}  (????).

\bibitem[{\citenamefont{Cox and Sleight}(1979)}]{Cox1979}
\bibinfo{author}{\bibfnamefont{D.~E.} \bibnamefont{Cox}} \bibnamefont{and}
  \bibinfo{author}{\bibfnamefont{A.~W.} \bibnamefont{Sleight}},
  \bibinfo{journal}{Acta Crystallographica Section B}
  \textbf{\bibinfo{volume}{35}}, \bibinfo{pages}{1} (\bibinfo{year}{1979}),
  \urlprefix\url{https://doi.org/10.1107/S0567740879002417}.

\bibitem[{\citenamefont{Rice and Sneddon}(1981)}]{Rice}
\bibinfo{author}{\bibfnamefont{T.~M.} \bibnamefont{Rice}} \bibnamefont{and}
  \bibinfo{author}{\bibfnamefont{L.}~\bibnamefont{Sneddon}},
  \bibinfo{journal}{Phys. Rev. Lett.} \textbf{\bibinfo{volume}{47}},
  \bibinfo{pages}{689} (\bibinfo{year}{1981}),
  \urlprefix\url{https://link.aps.org/doi/10.1103/PhysRevLett.47.689}.

\bibitem[{\citenamefont{Kim et~al.}(2015)\citenamefont{Kim, Neumann, Kim, Le,
  Kang, and Noh}}]{KimPRL2015}
\bibinfo{author}{\bibfnamefont{G.}~\bibnamefont{Kim}},
  \bibinfo{author}{\bibfnamefont{M.}~\bibnamefont{Neumann}},
  \bibinfo{author}{\bibfnamefont{M.}~\bibnamefont{Kim}},
  \bibinfo{author}{\bibfnamefont{M.~D.} \bibnamefont{Le}},
  \bibinfo{author}{\bibfnamefont{T.~D.} \bibnamefont{Kang}}, \bibnamefont{and}
  \bibinfo{author}{\bibfnamefont{T.~W.} \bibnamefont{Noh}},
  \bibinfo{journal}{Phys. Rev. Lett.} \textbf{\bibinfo{volume}{115}},
  \bibinfo{pages}{226402} (\bibinfo{year}{2015}),
  \urlprefix\url{https://link.aps.org/doi/10.1103/PhysRevLett.115.226402}.

\bibitem[{\citenamefont{Trivedi et~al.}(1996)\citenamefont{Trivedi, Scalettar,
  and Randeria}}]{TrivediPRB}
\bibinfo{author}{\bibfnamefont{N.}~\bibnamefont{Trivedi}},
  \bibinfo{author}{\bibfnamefont{R.~T.} \bibnamefont{Scalettar}},
  \bibnamefont{and} \bibinfo{author}{\bibfnamefont{M.}~\bibnamefont{Randeria}},
  \bibinfo{journal}{Phys. Rev. B} \textbf{\bibinfo{volume}{54}},
  \bibinfo{pages}{R3756} (\bibinfo{year}{1996}),
  \urlprefix\url{https://link.aps.org/doi/10.1103/PhysRevB.54.R3756}.

\bibitem[{\citenamefont{Naamneh et~al.}(2018)\citenamefont{Naamneh, Yan,
  Jandke, Ma, Risti\'{c}, Teyssier, Stucky, Marel, Gawryluk, Shang
  et~al.}}]{Naamneh}
\bibinfo{author}{\bibfnamefont{M.}~\bibnamefont{Naamneh}},
  \bibinfo{author}{\bibfnamefont{M.}~\bibnamefont{Yan}},
  \bibinfo{author}{\bibfnamefont{J.}~\bibnamefont{Jandke}},
  \bibinfo{author}{\bibfnamefont{J.}~\bibnamefont{Ma}},
  \bibinfo{author}{\bibfnamefont{J.}~\bibnamefont{Risti\'{c}}},
  \bibinfo{author}{\bibfnamefont{J.}~\bibnamefont{Teyssier}},
  \bibinfo{author}{\bibfnamefont{A.}~\bibnamefont{Stucky}},
  \bibinfo{author}{\bibfnamefont{D.}~\bibnamefont{Marel}},
  \bibinfo{author}{\bibfnamefont{D.~J.} \bibnamefont{Gawryluk}},
  \bibinfo{author}{\bibfnamefont{T.}~\bibnamefont{Shang}},
  \bibnamefont{et~al.}, \bibinfo{journal}{arXiv: 1808.06135}
  (\bibinfo{year}{2018}).

\bibitem[{\citenamefont{Bischofs et~al.}(2002)\citenamefont{Bischofs, Allen,
  Kostur, and Bhargava}}]{IlkaPRB2002}
\bibinfo{author}{\bibfnamefont{I.~B.} \bibnamefont{Bischofs}},
  \bibinfo{author}{\bibfnamefont{P.~B.} \bibnamefont{Allen}},
  \bibinfo{author}{\bibfnamefont{V.~N.} \bibnamefont{Kostur}},
  \bibnamefont{and} \bibinfo{author}{\bibfnamefont{R.}~\bibnamefont{Bhargava}},
  \bibinfo{journal}{Phys. Rev. B} \textbf{\bibinfo{volume}{66}},
  \bibinfo{pages}{174108} (\bibinfo{year}{2002}),
  \urlprefix\url{https://link.aps.org/doi/10.1103/PhysRevB.66.174108}.

\bibitem[{\citenamefont{Giraldo-Gallo et~al.}(2015)\citenamefont{Giraldo-Gallo,
  Zhang, Parra, Manoharan, Beasley, Geballe, Kramer, and Fisher}}]{Gallo}
\bibinfo{author}{\bibfnamefont{P.}~\bibnamefont{Giraldo-Gallo}},
  \bibinfo{author}{\bibfnamefont{Y.}~\bibnamefont{Zhang}},
  \bibinfo{author}{\bibfnamefont{C.}~\bibnamefont{Parra}},
  \bibinfo{author}{\bibfnamefont{H.~C.} \bibnamefont{Manoharan}},
  \bibinfo{author}{\bibfnamefont{M.~R.} \bibnamefont{Beasley}},
  \bibinfo{author}{\bibfnamefont{T.~H.} \bibnamefont{Geballe}},
  \bibinfo{author}{\bibfnamefont{M.~J.} \bibnamefont{Kramer}},
  \bibnamefont{and} \bibinfo{author}{\bibfnamefont{I.~R.}
  \bibnamefont{Fisher}}, \bibinfo{journal}{Nature Communications}
  \textbf{\bibinfo{volume}{6}}, \bibinfo{pages}{8231} (\bibinfo{year}{2015}),
  \urlprefix\url{https://www.nature.com/articles/ncomms9231}.

\bibitem[{\citenamefont{Climent-Pascual
  et~al.}(2011)\citenamefont{Climent-Pascual, Ni, Jia, Huang, and
  Cava}}]{PhysRevB.83.174512}
\bibinfo{author}{\bibfnamefont{E.}~\bibnamefont{Climent-Pascual}},
  \bibinfo{author}{\bibfnamefont{N.}~\bibnamefont{Ni}},
  \bibinfo{author}{\bibfnamefont{S.}~\bibnamefont{Jia}},
  \bibinfo{author}{\bibfnamefont{Q.}~\bibnamefont{Huang}}, \bibnamefont{and}
  \bibinfo{author}{\bibfnamefont{R.~J.} \bibnamefont{Cava}},
  \bibinfo{journal}{Phys. Rev. B} \textbf{\bibinfo{volume}{83}},
  \bibinfo{pages}{174512} (\bibinfo{year}{2011}),
  \urlprefix\url{https://link.aps.org/doi/10.1103/PhysRevB.83.174512}.

\bibitem[{\citenamefont{Tajima et~al.}(1985)\citenamefont{Tajima, Uchida,
  Masaki, Takagi, Kitazawa, Tanaka, and Katsui}}]{PhysRevB.32.6302}
\bibinfo{author}{\bibfnamefont{S.}~\bibnamefont{Tajima}},
  \bibinfo{author}{\bibfnamefont{S.}~\bibnamefont{Uchida}},
  \bibinfo{author}{\bibfnamefont{A.}~\bibnamefont{Masaki}},
  \bibinfo{author}{\bibfnamefont{H.}~\bibnamefont{Takagi}},
  \bibinfo{author}{\bibfnamefont{K.}~\bibnamefont{Kitazawa}},
  \bibinfo{author}{\bibfnamefont{S.}~\bibnamefont{Tanaka}}, \bibnamefont{and}
  \bibinfo{author}{\bibfnamefont{A.}~\bibnamefont{Katsui}},
  \bibinfo{journal}{Phys. Rev. B} \textbf{\bibinfo{volume}{32}},
  \bibinfo{pages}{6302} (\bibinfo{year}{1985}),
  \urlprefix\url{https://link.aps.org/doi/10.1103/PhysRevB.32.6302}.

\bibitem[{\citenamefont{Nagata et~al.}(1999)\citenamefont{Nagata, Mishiro,
  Uchida, Ohtsuka, and Samata}}]{Nagata}
\bibinfo{author}{\bibfnamefont{Y.}~\bibnamefont{Nagata}},
  \bibinfo{author}{\bibfnamefont{A.}~\bibnamefont{Mishiro}},
  \bibinfo{author}{\bibfnamefont{T.}~\bibnamefont{Uchida}},
  \bibinfo{author}{\bibfnamefont{M.}~\bibnamefont{Ohtsuka}}, \bibnamefont{and}
  \bibinfo{author}{\bibfnamefont{H.}~\bibnamefont{Samata}},
  \bibinfo{journal}{Journal of Physics and Chemistry of Solids}
  \textbf{\bibinfo{volume}{60}}, \bibinfo{pages}{1933} (\bibinfo{year}{1999}),
  \urlprefix\url{https://www.sciencedirect.com/science/article/pii/S0022369799002176}.

\end{thebibliography}
\clearpage
\newpage
\widetext
\begin{figure*}[t] 
\center\includegraphics[page=1,width=\textwidth]{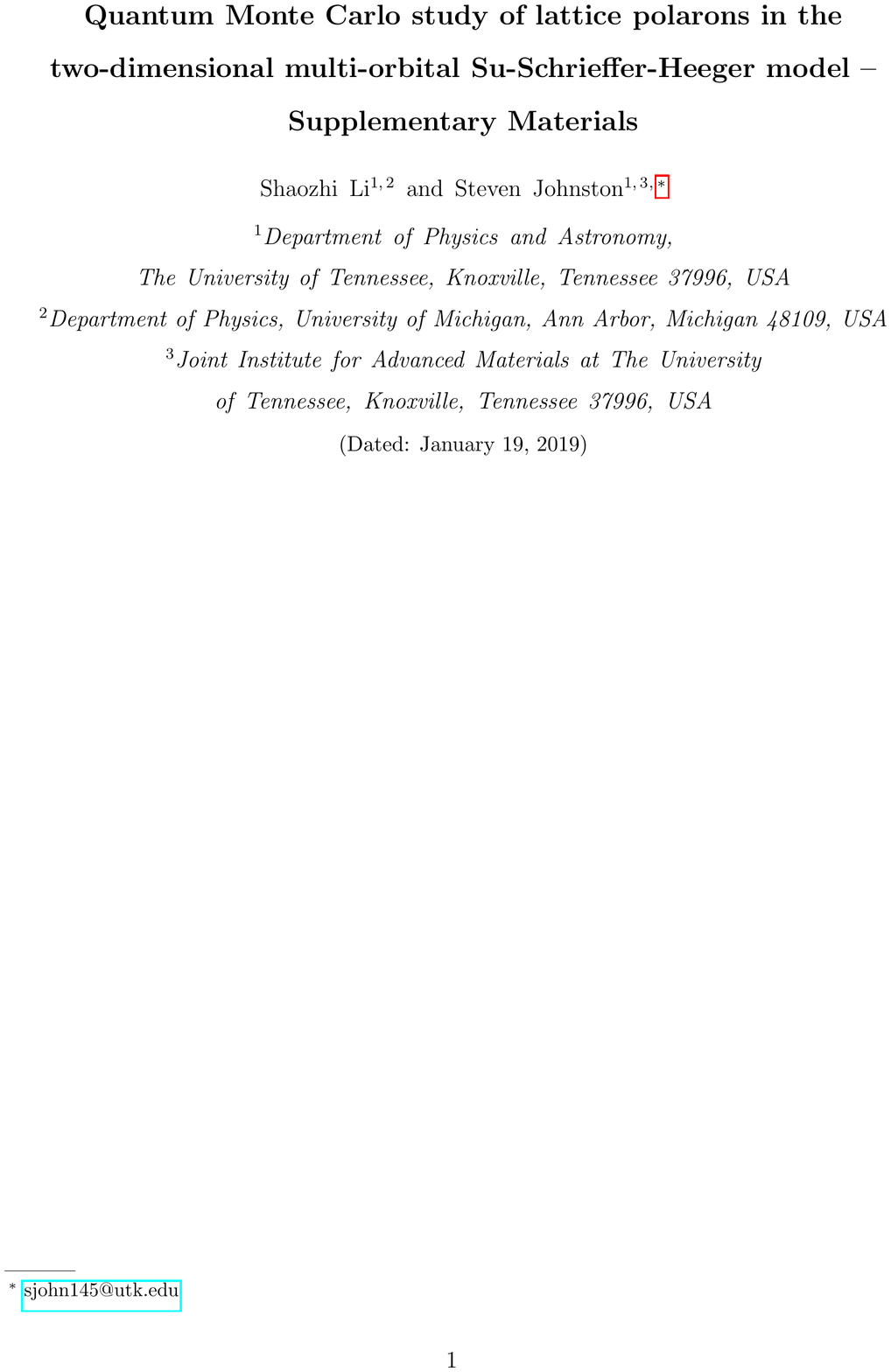}
\end{figure*}
\newpage
\begin{figure*}[t] 
\center\includegraphics[page=2,width=\textwidth]{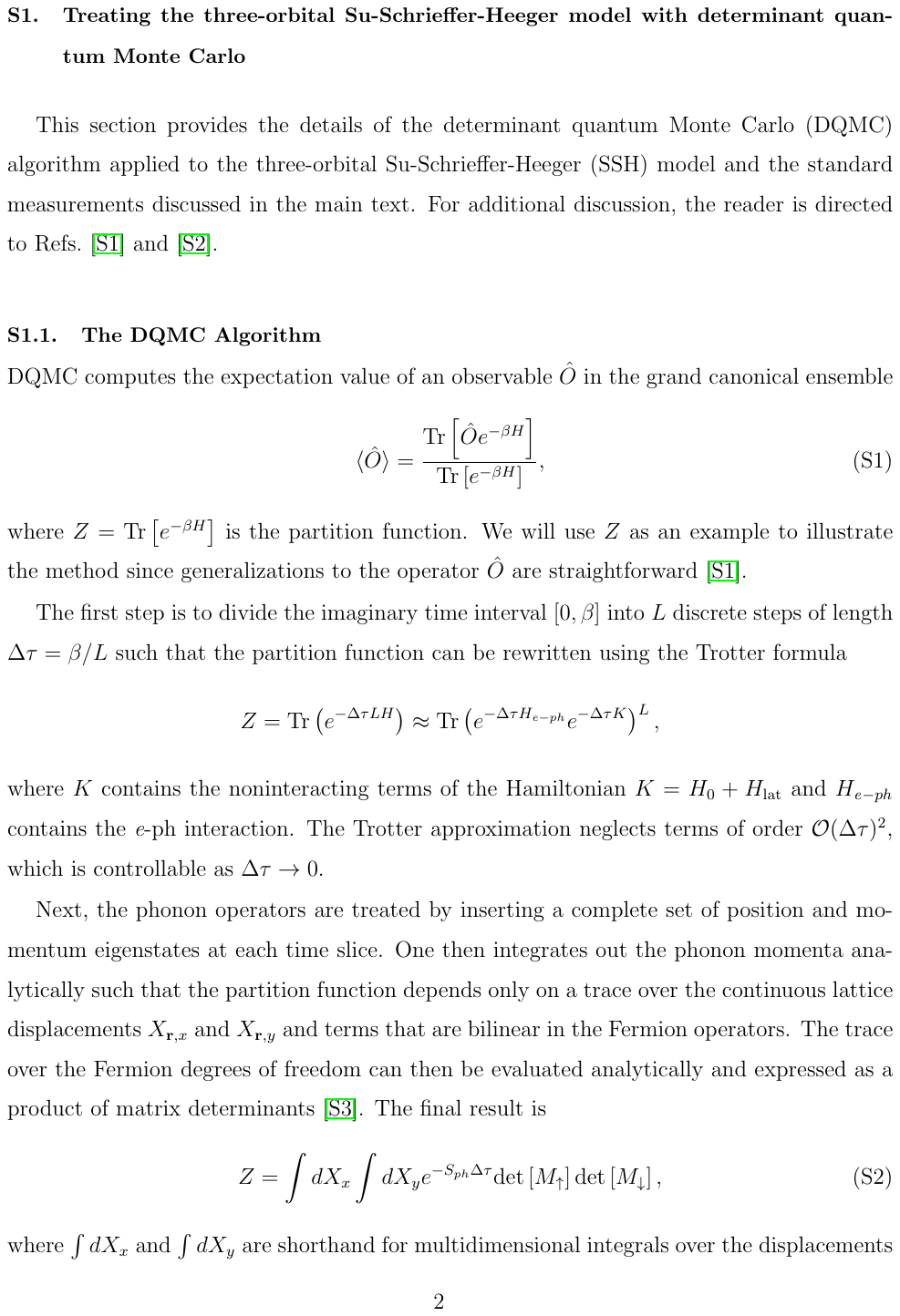}
\end{figure*}
\newpage
\begin{figure*}[t] 
\center\includegraphics[page=3,width=\textwidth]{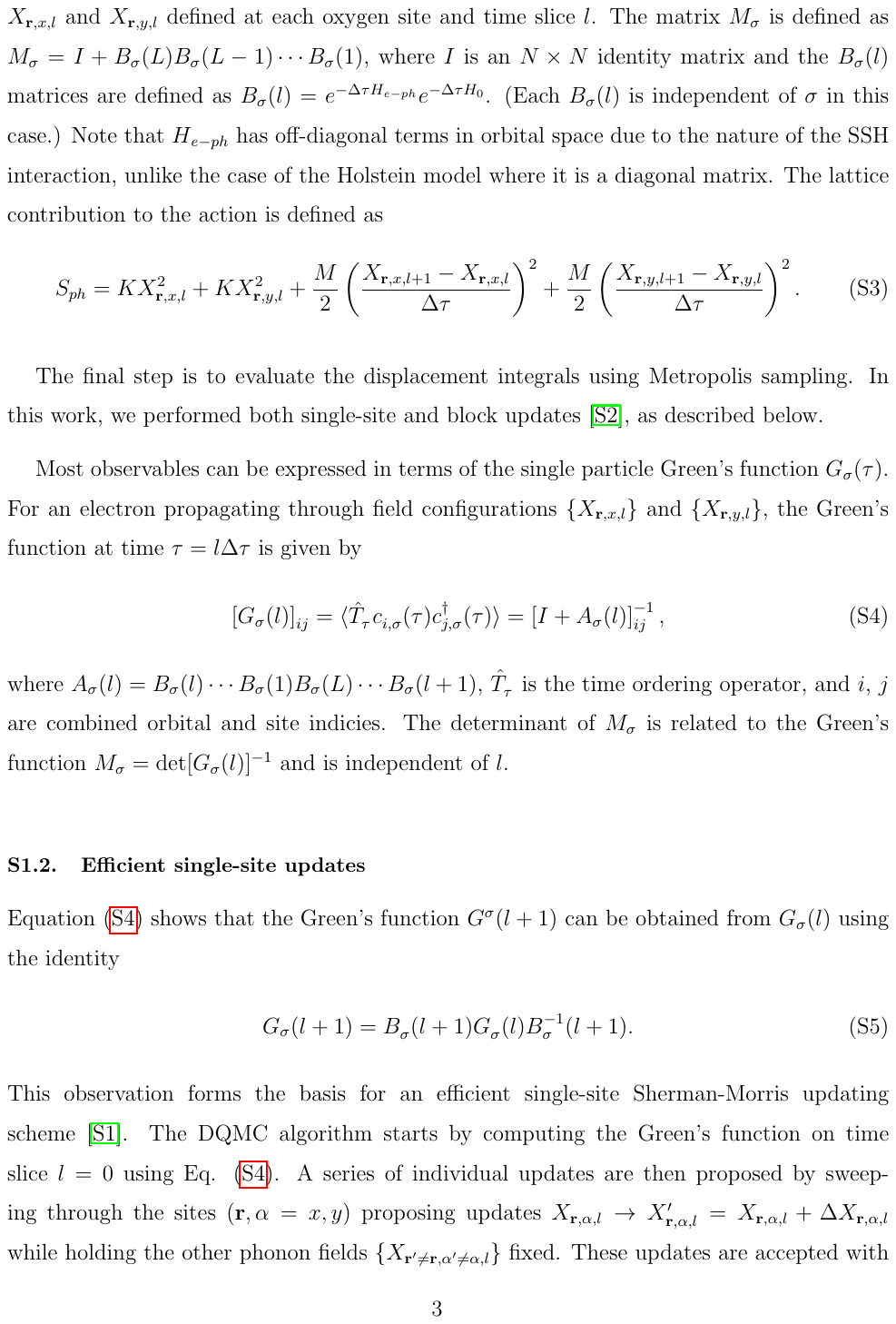}
\end{figure*}
\newpage
\begin{figure*}[t] 
\center\includegraphics[page=4,width=\textwidth]{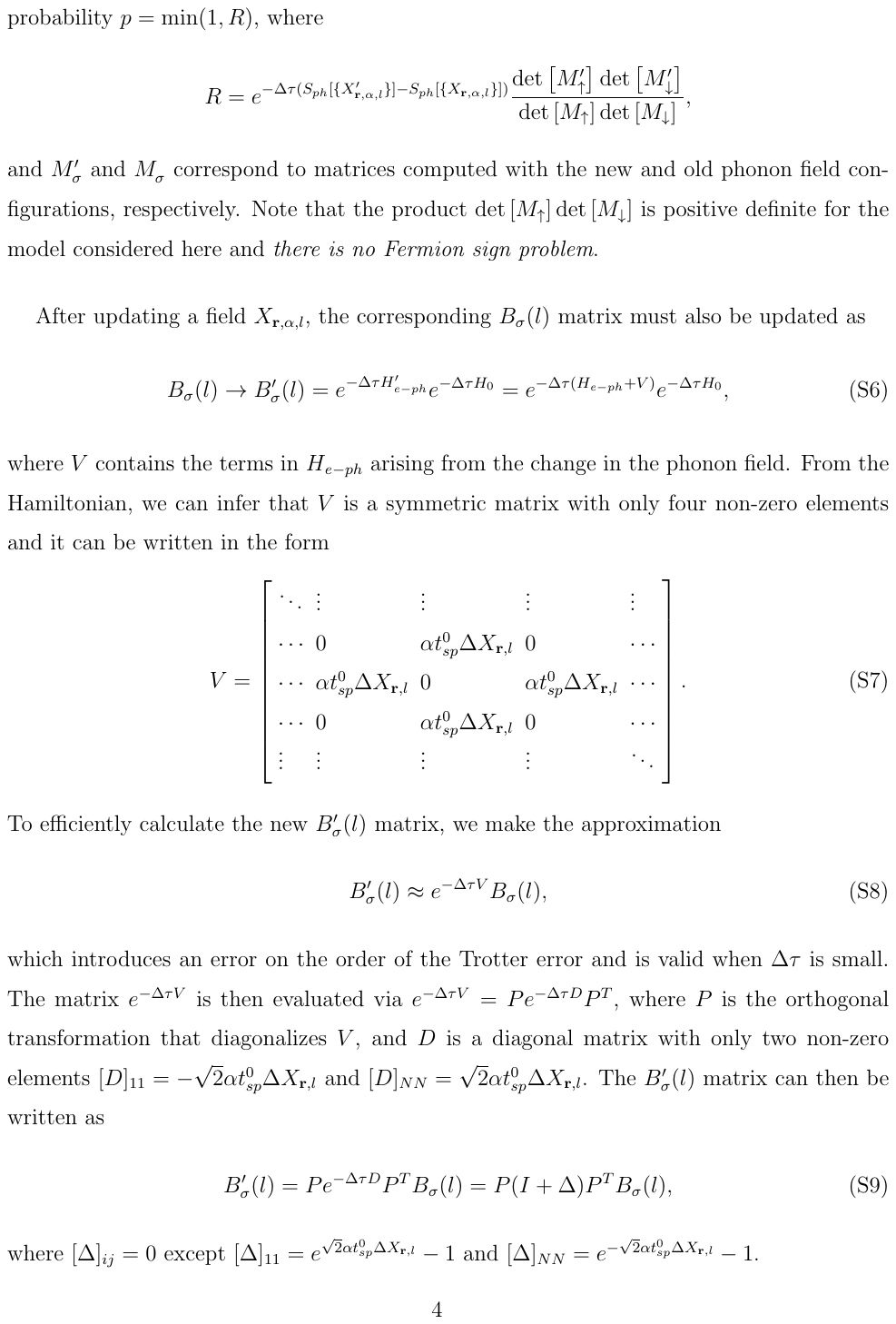}
\end{figure*}
\newpage
\begin{figure*}[t] 
\center\includegraphics[page=5,width=\textwidth]{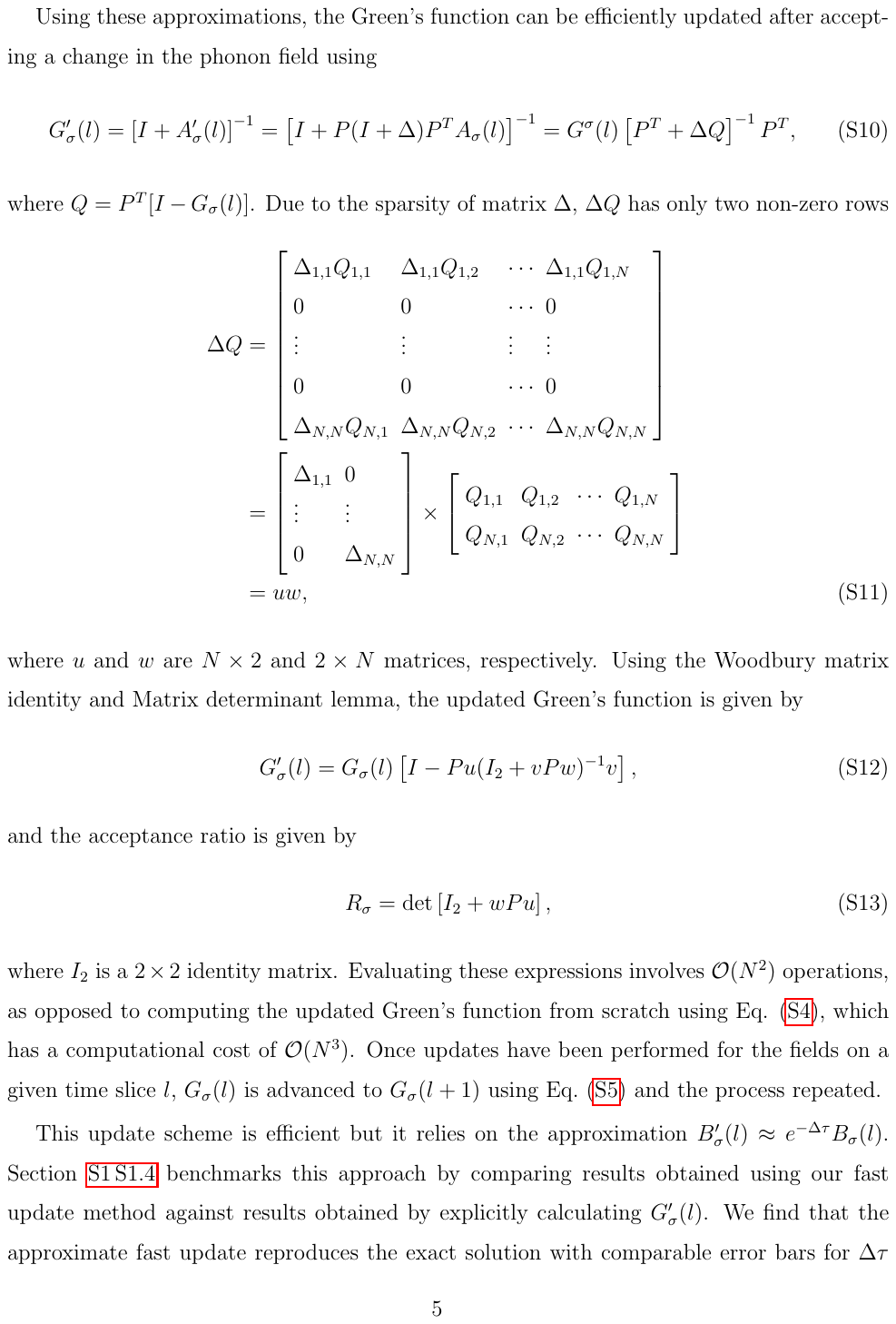}
\end{figure*}
\newpage
\begin{figure*}[t] 
\center\includegraphics[page=6,width=\textwidth]{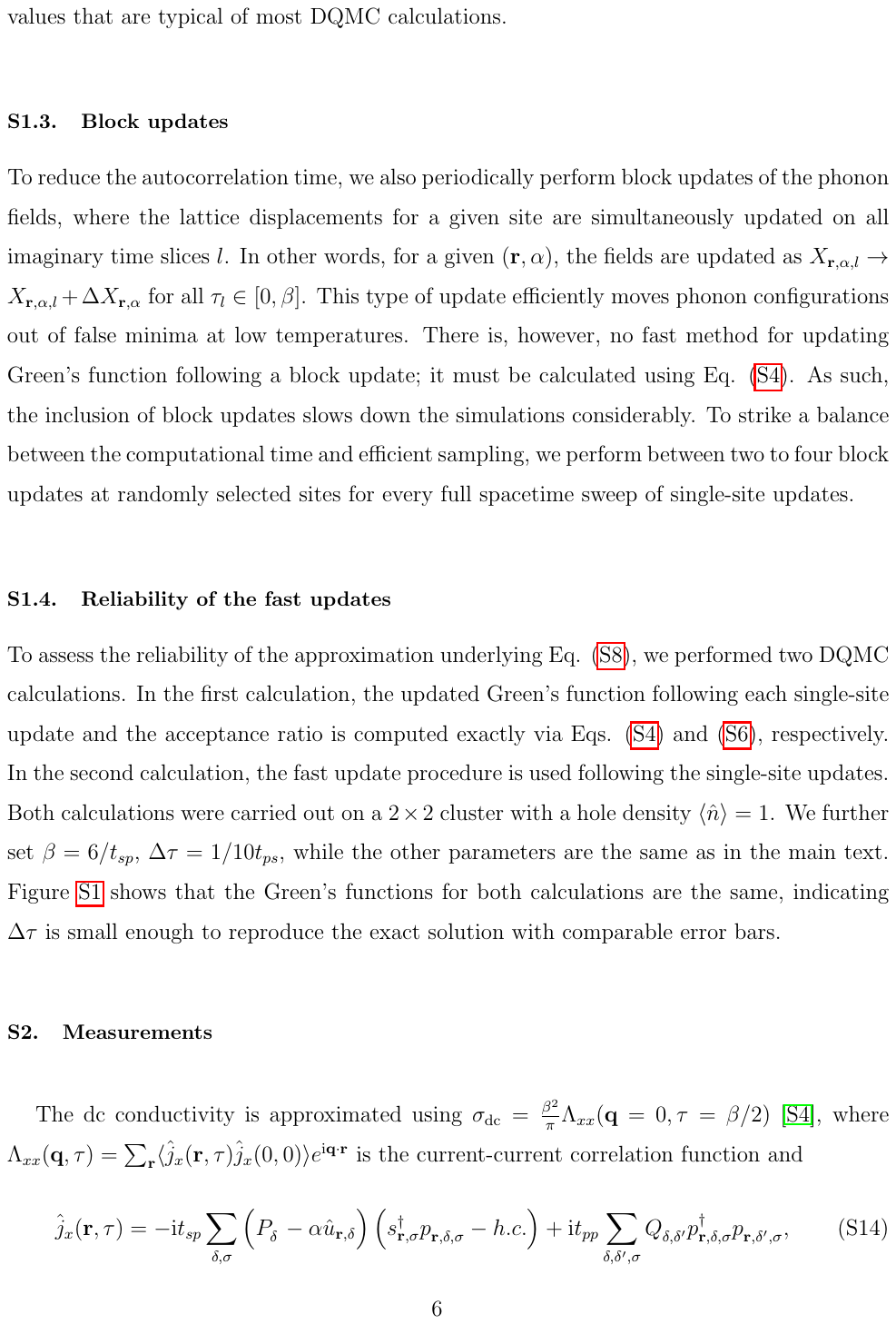}
\end{figure*}
\newpage
\begin{figure*}[t] 
\center\includegraphics[page=7,width=\textwidth]{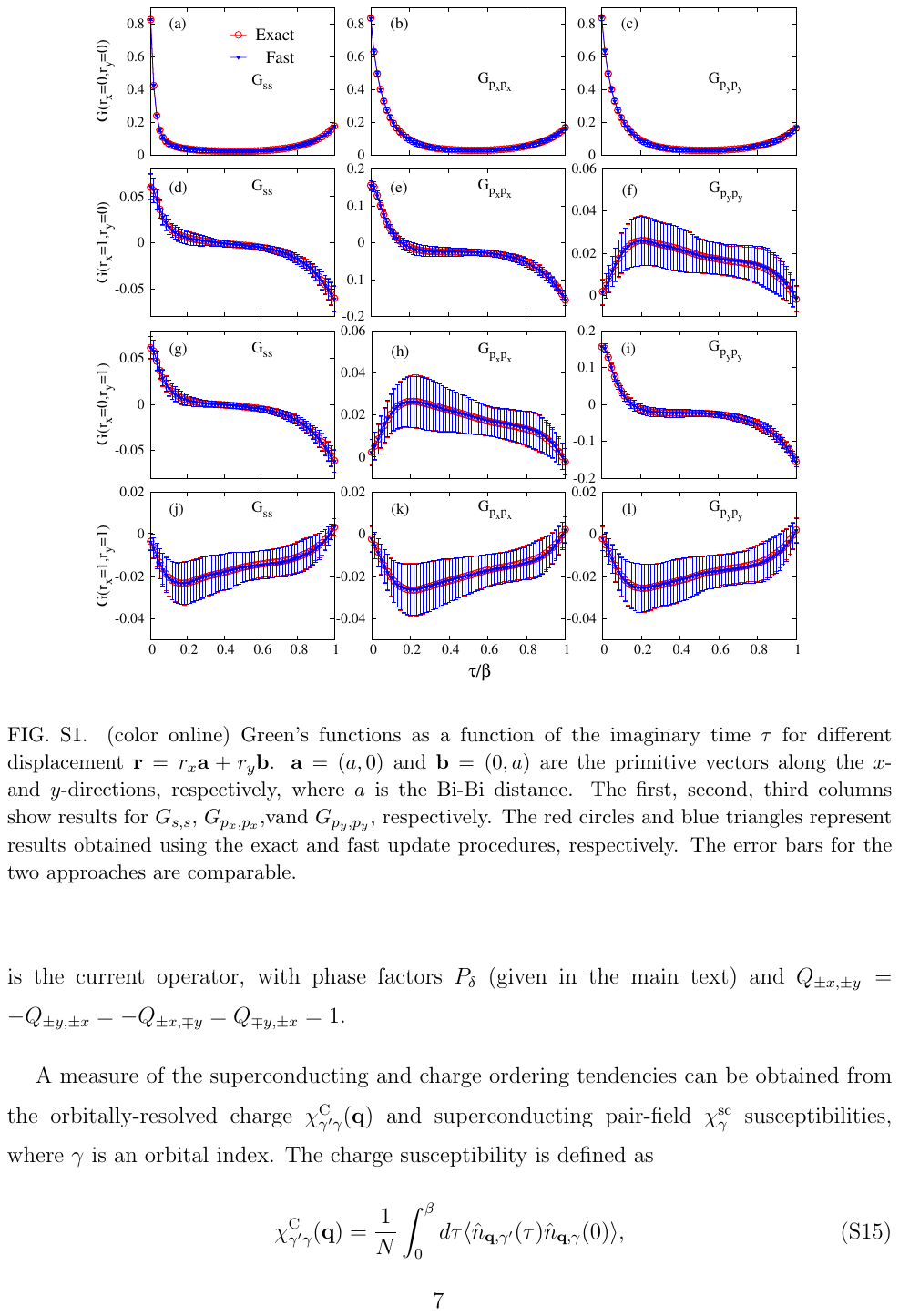}
\end{figure*}
\newpage
\begin{figure*}[t] 
\center\includegraphics[page=8,width=\textwidth]{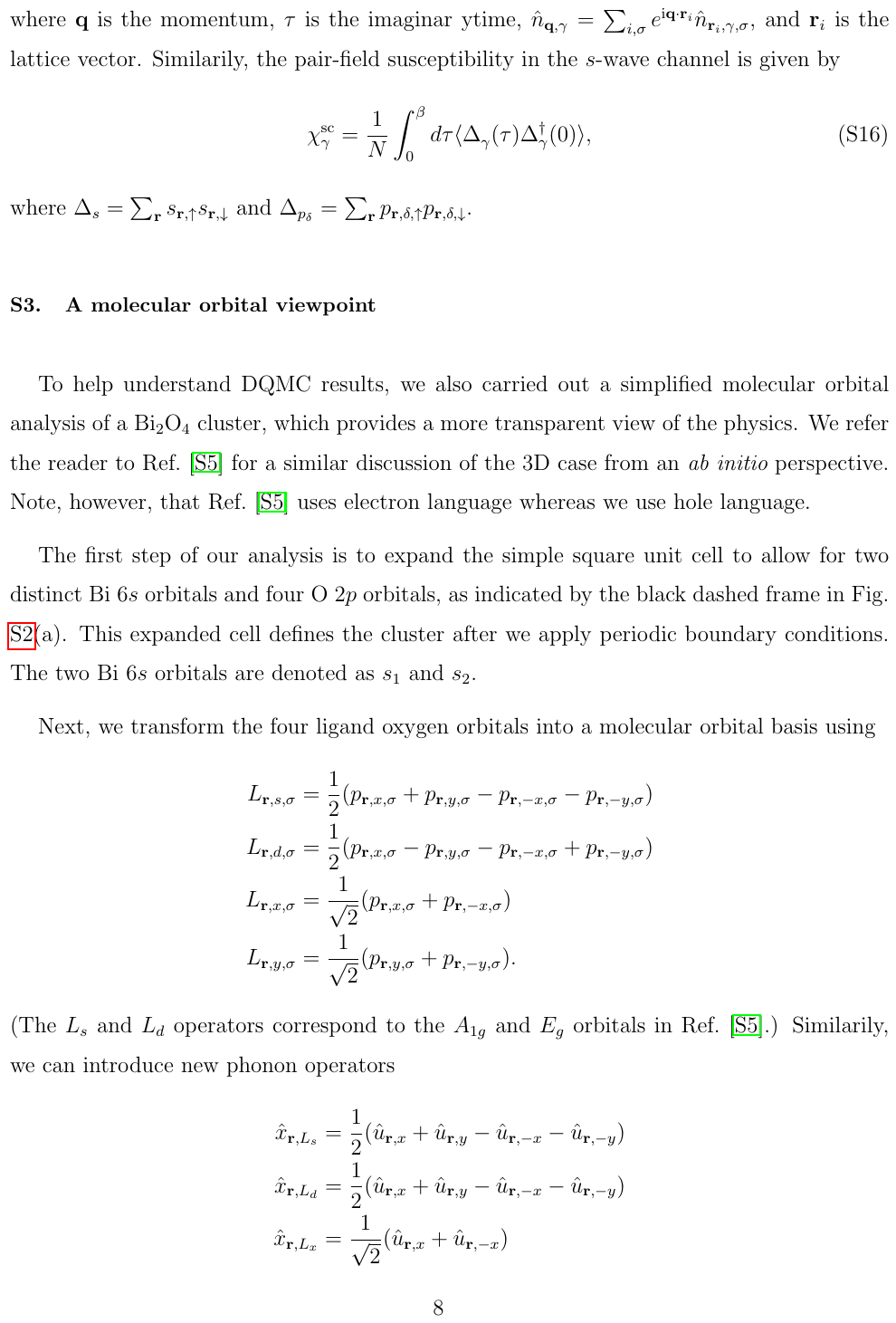}
\end{figure*}
\newpage
\begin{figure*}[t] 
\center\includegraphics[page=9,width=\textwidth]{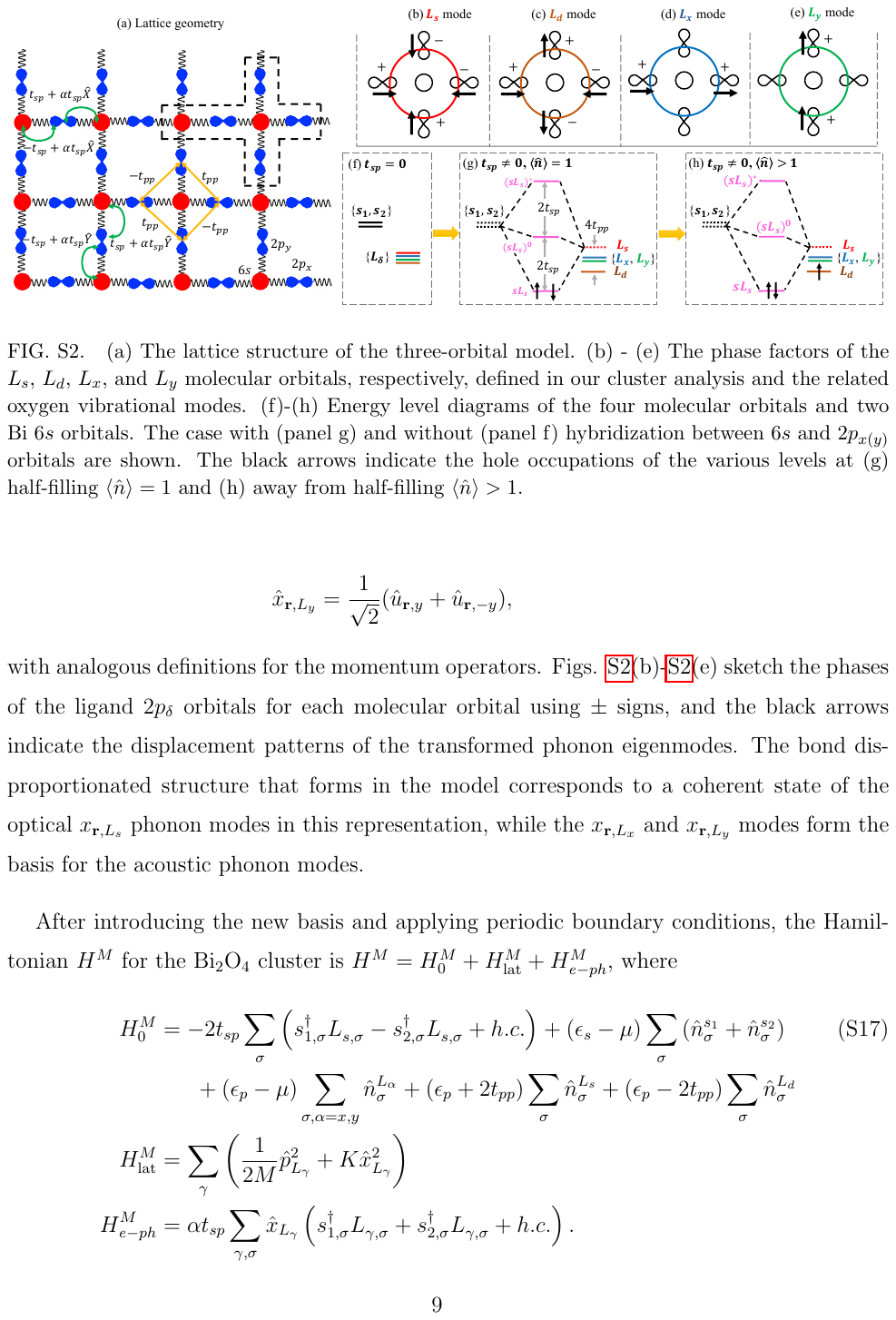}
\end{figure*}
\newpage
\begin{figure*}[t] 
\center\includegraphics[page=10,width=\textwidth]{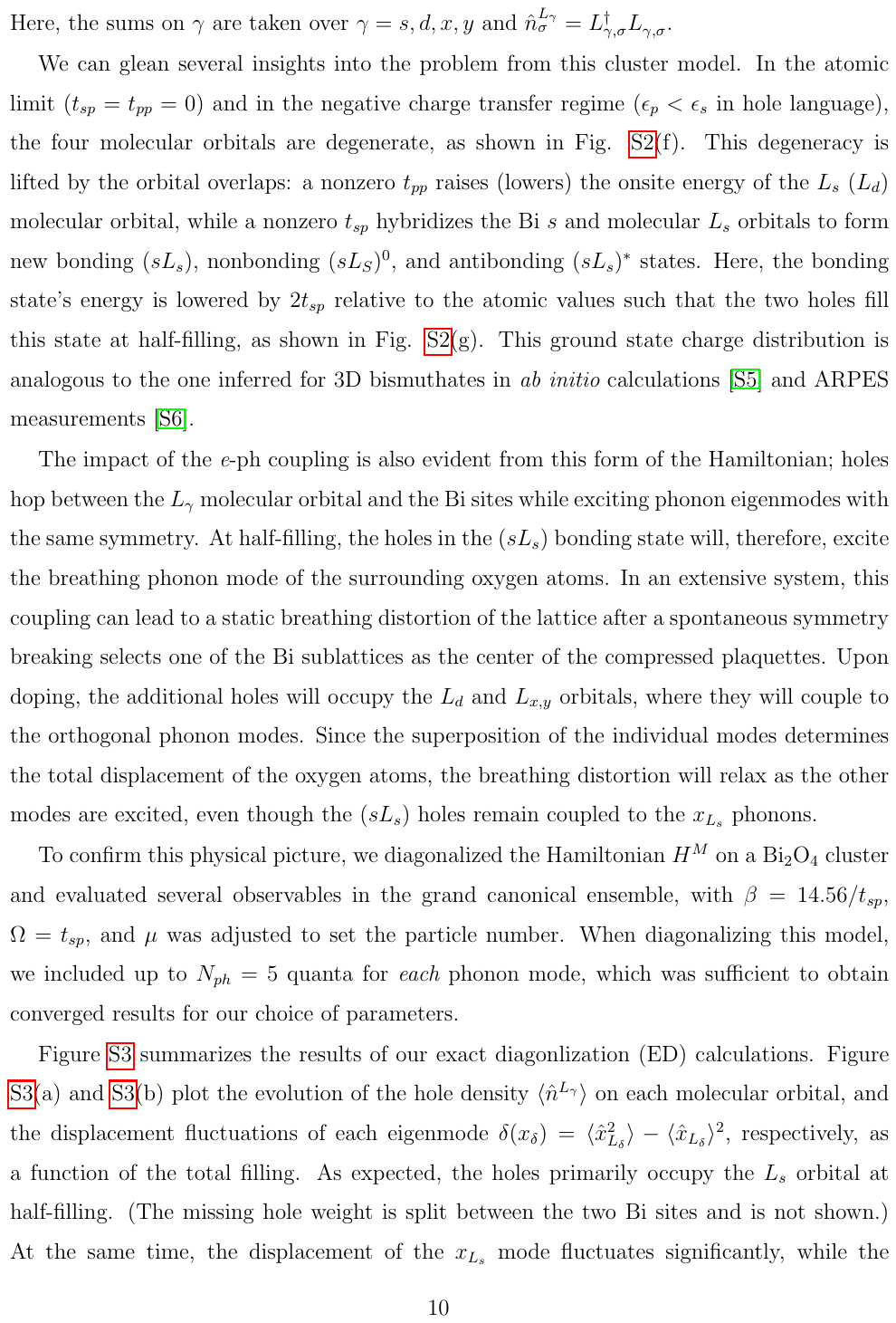}
\end{figure*}
\newpage
\begin{figure*}[t] 
\center\includegraphics[page=11,width=\textwidth]{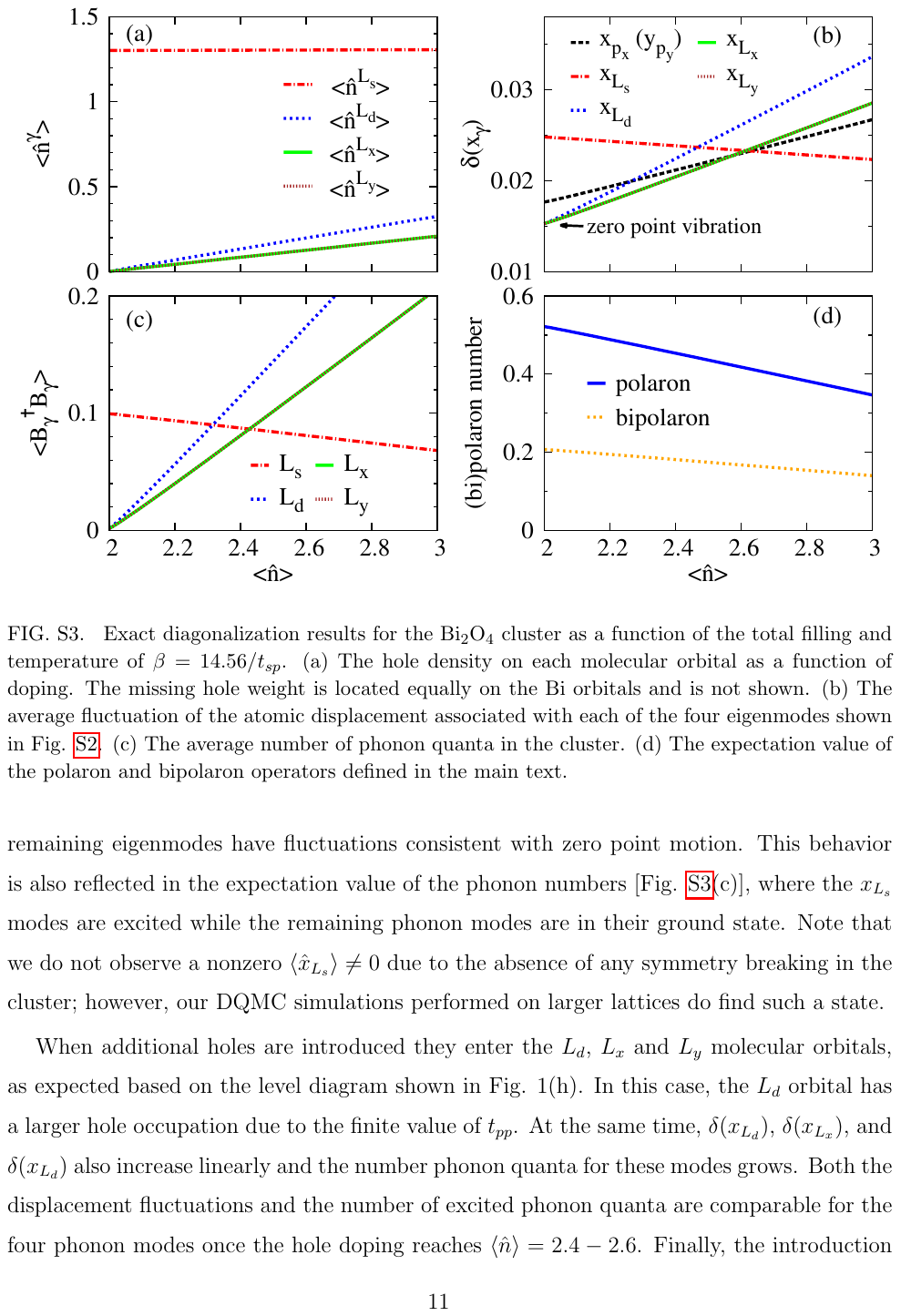}
\end{figure*}
\newpage
\begin{figure*}[t] 
\center\includegraphics[page=12,width=\textwidth]{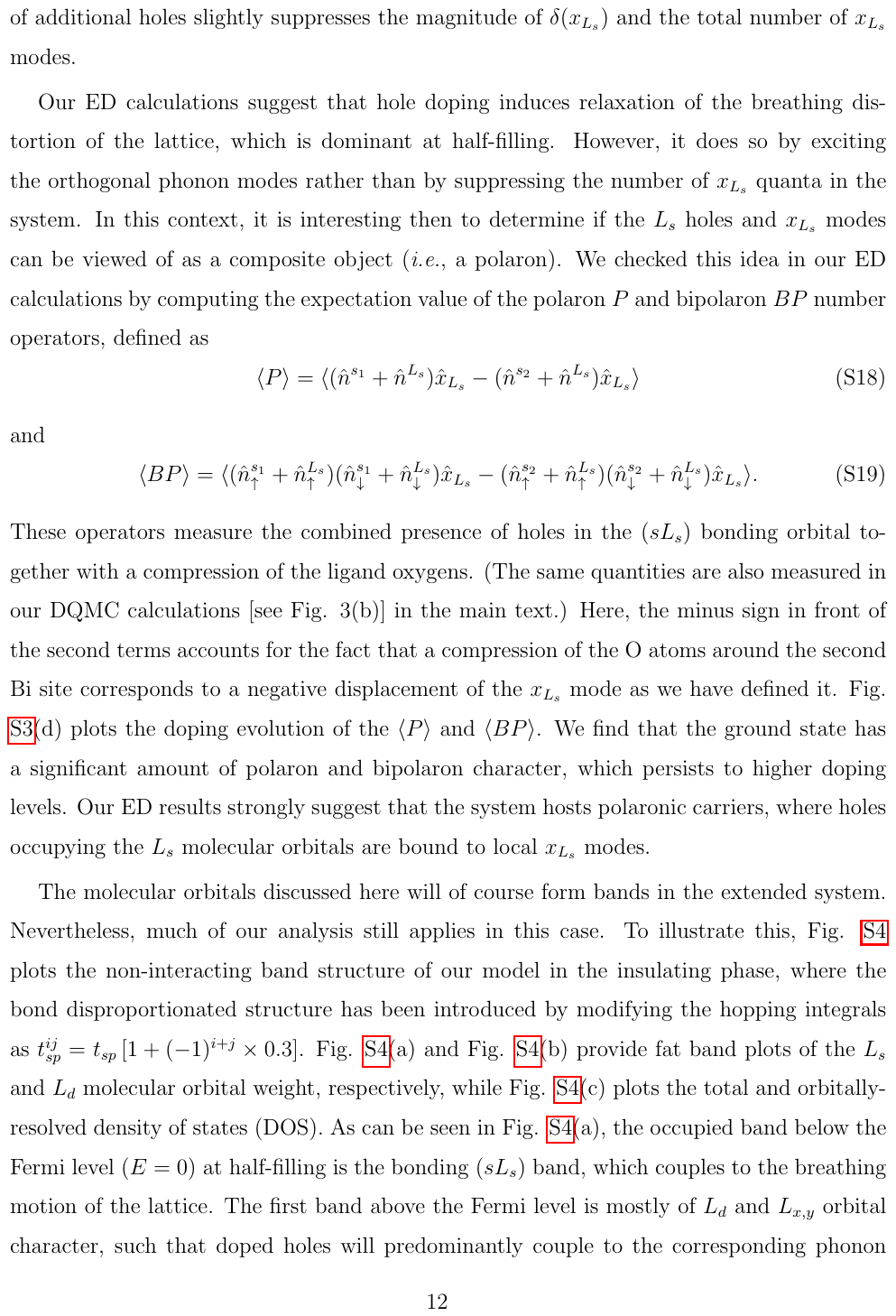}
\end{figure*}
\newpage
\begin{figure*}[t] 
\center\includegraphics[page=13,width=\textwidth]{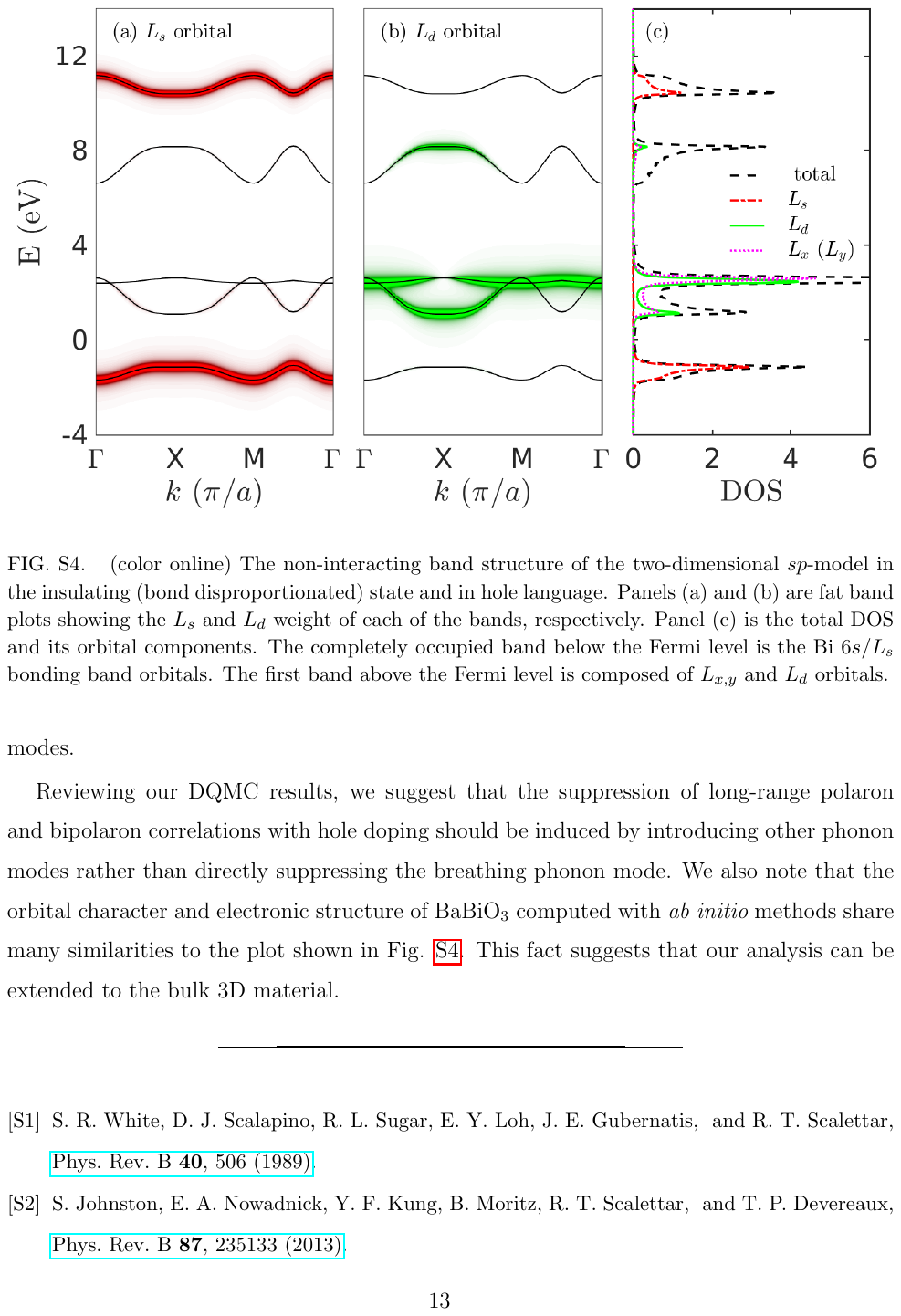}
\end{figure*}
\begin{figure*}[t] 
\center\includegraphics[page=14,width=\textwidth]{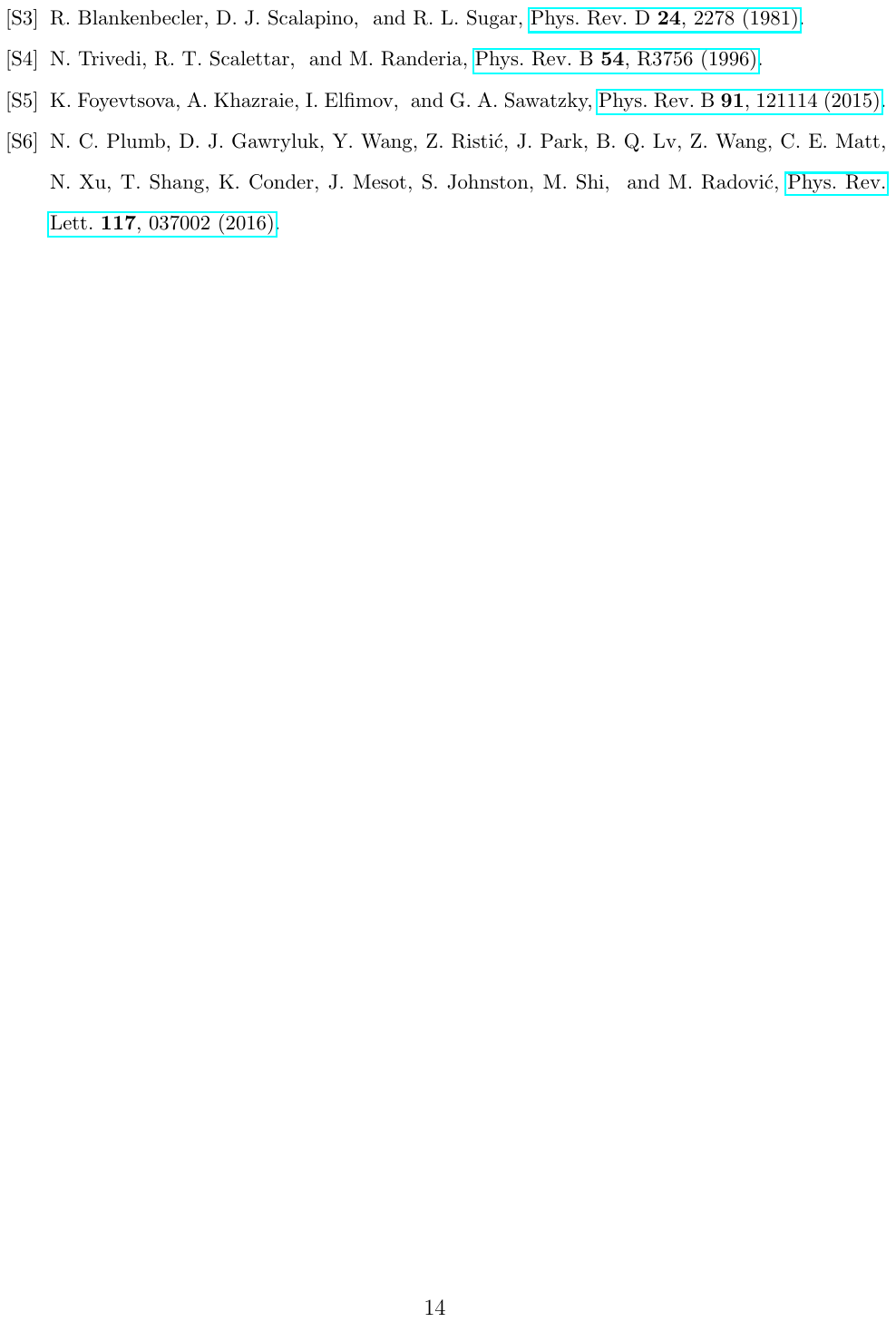}
\end{figure*}

\end{document}